\documentclass[aps,pra,superscriptaddress,notitlepage,showpacs,showkeys,twocolumn
]{revtex4-1}
\usepackage{graphicx,amsmath,amsfonts,amssymb,upgreek,txfonts,color}
\usepackage[colorlinks,linkcolor=blue,citecolor=blue,urlcolor=blue,breaklinks=true]{hyperref}
\usepackage{color, colortbl}
\definecolor{lightgreen}{rgb}{0.88,1,1}
\usepackage{textcomp}
\usepackage{tikz}
\usepackage{adjustbox}
\usepackage{setspace}
\usepackage{cleveref}
\usepackage{graphicx}
\usepackage[utf8]{inputenc}

\begin{document}

\title{ Optomechanically induced transparency, absorption, and conversion between slow and fast light  in a generalized cross-Kerr optomechanical circuit}

\author{S. \surname{Bayati} }
\email{phsnbayati@gmail.com}
\address{Department of Physics, University of Isfahan, Hezar Jerib, 81746-73441, Isfahan, Iran}

\author{M.H. \surname{Naderi}}
\email{mhnaderi@sci.ui.ac.ir}
\address{Department of Physics, University of Isfahan, Hezar Jerib, 81746-73441, Isfahan, Iran}
\address{Quantum Optics Group, Department of Physics, University of Isfahan, Hezar Jerib, 81746-73441, Isfahan, Iran}

\begin{abstract}
 In this paper, we propose and explore an experimentally viable scheme to realize tunable optomechanically induced transparency (\text{OMIT}) and optomechanically induced absorption (\text{OMIA}) phenomena in a hybrid microwave-optomechanical circuit in which two single-Cooper-pair transistors (\text{SCPTs}) are coupled to a common microwave $LC$ resonator and two independent micromechanical resonators. We show that under special conditions such a system can be equivalently modeled as a two-mechanical-modes optomechanical cavity in which, besides the standard radiation-pressure coupling, the cavity mode interacts with the mechanical modes through the cross-Kerr (\text{CK}), a higher-order generalized CK, and a three-mode \text{CK} type of coupling. Furthermore, there is an induced \text{CK} coupling between the two-mechanical modes. Assuming that the cavity mode is simultaneously driven by a strong control field and a weak probe field, we analyze the response of the output probe field affected by the above-mentioned nonlinear couplings. In particular, our results reveal that the higher-order nonlinear CK and the three-mode \text{CK} couplings have remarkable impact on the characteristics of the \text{OMIT} and \text{OMIA} phenomena. Moreover, we find that these nonlinear couplings can give rise to the occurrence of the gain in the absorption profile and contribute to the amplification of the output probe field in specific frequency regions. We also show that the system offers tunable switching between slow and fast light behaviors. The proposed hybrid optomechanical circuit may find potential applications in light propagation, quantum sensing of physical quantities, and information processing.
\end{abstract}

\maketitle

\section{\label{sec1} Introduction}

The coherent interaction between laser radiation and multilevel atoms leads to interesting phenomena, most notably electromagnetically induced transparency (EIT) \cite{1,2,3,4} and electromagnetically induced absorption (EIA) \cite{5n,5nn}. EIT stems from the destructive quantum interference between two different absorption pathways in a multi-level atomic system which is simultaneously driven by a probe laser and a coupling laser. This effect effectively enables an opaque medium to become transparent near the resonance frequency, resulting in significant changes in the absorption and dispersion properties of the medium. On the other hand, EIA is the result of constructive interference between different pathways. Since the first observation of EIT in cold atomic ensembles \cite{1}, atomic EIT has been successfully utilized for a variety of applications such as lasing without population inversion \cite{6n}, slowing and stopping of pulses of light \cite{7n,8n,9n}, storing and retrieving single photons in hot and cold atomic gases \cite{10n, 11n, 12n}, and stimulated Raman adiabatic passage \cite{13n}. Moreover, atomic EIT has been applied for the implementation of quantum switching \cite{14n}, controlled generation of single photons \cite{15n}, and coherent manipulation and entanglement of collective excitations in mesoscopic ensembles of cold atoms \cite{16n}. Inspired by the significant success of atomic EIT studies, considerable theoretical and experimental investigations have been devoted to realize analogs of EIT in various physical settings, including quantum wells \cite{17n}, quantum dots \cite{18n}, metamaterials \cite{19n, 20n}, nitrogen-vacancy centers \cite{21n}, superconducting qubits in a microwave resonator \cite{22n}, ion Coulomb crystals placed inside an optical cavity \cite{23n}, cavity magnomechanical systems \cite{24n,25n,26n}, nano-electromechanical systems \cite{27n,28n,29n}, and cavity optomechanical systems \cite{4,31n,32n,33n}. 
\\

The analog of the EIT effect in cavity optomechanics, the so-called optomechanically induced transparency (OMIT), is caused by radiation pressure to couple light to mechanical resonator modes \cite{31n}. The underlying physical mechanism is related to the destructive interference between  a weak probe field and the anti-Stokes scattering of light from a strong red-detuned control field induced by the mechanical oscillation. Due to the interference, the transmission of the probe field exhibits a narrow peak around the cavity resonance, the width of which is determined by the mechanical relaxation rate. In addition to OMIT, a similar phenomenon to the EIA \cite{34n} in atomic systems, termed as optomechanically induced absorption (OMIA) \cite{32n,35n,36n} can be realized by setting the detuning of the control field to the blue side of the optomechanical cavity resonance. As a counterpart of OMIT, the effect of OMIA arises due to the constructive interference between the probe field and the Stokes sideband field, leading to a robust absorption peak within the transparency window. Thanks to the development of nanophotonics along with the improvement of nano-fabrication techniques, OMIT has been found to manifest in a wide range of experimental settings including, among others, a microwave circuit \cite{37n}, a microtoroid resonator \cite{31n}, a membrane-in-the-middle cavity \cite{38n}, an optical whispering gallery mode coupled to two mechanical breathing modes in a silica microsphere \cite{39n}, a diamond optomechanical microdisk cavity \cite{40n}, and a nonlinear Kerr resonator \cite{41n}. In parallel, numerous studies show that OMIT can exhibit great potential to realize several important goals, such as precision measurement of coupling rate \cite{42n}, measuring a weak magnetic field \cite{43n}, precision measurement of electrical charges \cite{44n}, slow and fast light \cite{45n,46n}, storage of optical information in the long-lived mechanical oscillations \cite{47n}, and quantum information processing \cite{48n,49n}. Besides the study of conventional OMIT in the standard optomechanical systems comprised of one optical mode and one mechanical mode, a variety of generalized optomechanical schemes have been proposed to get more ﬂexible controllability of OMIT. Some examples include OMIT in hybridized cavity optomechanical systems \cite{46n,50n,51n,53n,54n,55n}, OMIT in  the vector cavity optomechanics \cite{56n},  nonreciprocal OMIT \cite{57n,58n,59n}, reversed OMIT \cite{60n,61n,62n}, double OMIT \cite{63n,8,64n}, multiple OMIT \cite{65n,66n}, OMIT in quadratically coupled optomechanical cavities \cite{67n, 68n}, OMIT in the nonlinear quantum regime \cite{69n,70n,71n}, and OMIT in higher-order sidebands \cite{72n}.
\\

Nonlinearity is an intrinsic feature of the photon-phonon optomechanical coupling. It stems from the fact that the length of the cavity depends upon the intensity of the field in an analogous way to the optical length of a Kerr medium \cite{73n,74n}. However, this coupling is relatively weak, which limits the observation of nontrivial quantum phenomena in the single-photon optomechanical coupling regime \cite{75n}.  So far, a variety of proposals have been theoretically designed and experimentally explored to conquer this challenge. The idea of coupling nonlinear optical devices such as Kerr medium and/or optical parametric amplifier (OPA) with an optomechanical cavity has been widely investigated with the purpose of enhancing quantum effects. The integration of these nonlinear devices with the optomechanical system can lead to intriguing phenomena, e.g., the normal mode splitting of the movable mirror and the output field \cite{76n}, improved ground-state cooling of a mechanical oscillator \cite{77n,78n,79n}, quantum synchronization of two mechanical oscillators \cite{80n}, enhanced photon blockade \cite{81n}, strong mechanical  squeezing \cite{82n,83n}, enhanced optomechanical entanglement \cite{78n,85n}, and enhanced sensing precision \cite{86n,87n,88n,89n}. In addition, several interesting studies have been carried out in recent years to investigate the optical-response properties of  nonlinear cavity optomechanical systems hybridized with a Kerr- type medium \cite{90n,91n,92n}, with an OPA \cite{93n,94n,95n}, or with a Kerr-down-conversion crystal \cite{8}.
\\

Besides the above-mentioned platforms, hybrid microwave optomechanical systems coupled to a nonlinear inductive element (single-Cooper-pair transistor (SCPT)) \cite{96n,97n}, a suspended carbon nanotube quantum dot \cite{98n}, a nonlinear capacitive element (Cooper-pair box) \cite{99n}, or a quantum two-level system (qubit) \cite{100n} have proven to be promising candidates for significant improvement of the optomechanical coupling. (We refer the reader to Ref. \cite{101n} for a comprehensive review of recent developments in the field of amplification of quantum light–matter coupling, particularly in cavity and circuit quantum electrodynamics and in cavity optomechanics.) In these hybrid systems not only the optomechanical coupling can be strengthened several orders of magnitude, but also an additional nonlinear cross-Kerr (\text{CK}) type coupling $g_{\text{CK}}\hat{n}_a \hat{n}_b$ between the cavity and the mechanical resonator, with respective number operators $\hat{n}_a$ and $\hat{n}_b$, is induced \cite{96n,99n}. Such a nonlinear coupling leads to a dispersive frequency shift in the mechanical (cavity) mode with linear dependence versus the photon (phonon) number, as well as an optimal cooling or heating of the mechanical oscillator \cite{103n}. It has been shown \cite{104n} that in the presence of \text{CK} coupling a significantly enhanced steady-state entanglement between the cavity and mechanical modes can be achieved, which is extremely robust against thermal phonon fluctuations. In addition, the \text{CK} nonlinearity can substantially enhance the cooling of a rotating mirror as well as its steady-state entanglement with a cavity mode in a Laguerre–Gaussian-cavity optomechanical system \cite{105n,106n}. As theoretically shown in Ref. \cite{107n}, the \text{CK} nonlinearity can also be employed to achieve simultaneous ground-state cooling of multiple degenerate mechanical resonators. In Ref. \cite{108n} the influence of the \text{CK} nonlinearity on the photon blockade, the magnitude of single-photon mechanical displacement, and the mechanical cat state generation in a superconducting quantum circuit has been investigated. In Ref. \cite{109n} the authors have investigated the impacts of the \text{CK} nonlinearity on the steady-state behavior of the mean phonon number as well as optical transparency in an optomechanical system driven by two-tone fields. The OMIT phenomenon influenced by the \text{CK} nonlinearity in a parity-time symmetric optomechanical system has also been studied \cite{110n}. Meanwhile, a number of theoretical protocols have been developed to amplify the optomechanical cross-Kerr interaction based on, for example, periodic modulation of the mechanical spring constant \cite{111n}, strong mechanical driving \cite{112n}, utilization of Josephson (quantum) capacitance of a Cooper-pair box \cite{99n}, and two-photon parametric driving \cite{114n}. 
\\

More recently, a theoretical scheme has been proposed \cite{115n} that makes it possible not only to achieve the strong optomechanical and \text{CK} photon-phonon couplings, but also to attain an additional higher-order nonlinear (generalized) \text{CK} type of coupling in a tripartite microwave-optomechanical system composed of a single SCPT, a microwave $LC$ resonator, and a single micromechanical resonator. This type of \text{CK} nonlinearity, which is linearly dependent on photon number while quadratically dependent on mechanical phonon number (i.e., proportional to $\hat{n}_a \hat{n}_b^2$) can be realized via adjusting the gate charge of the Cooper-pair transistor. Focusing on the few-photon optomechanical effects, the authors have found that the presence of both conventional \text{CK} and generalized \text{CK} nonlinearities leads to the enhancement of one- and two-photon blockades as well as photon-induced tunneling, provides the possibility of generating dissipation-robust multicomponent mechanical superposition states, and gives rise to the increase of the microwave-mechanical bipartite entanglement in the regime of  large red detuning.
\\

Inspired by those previous works, in this paper we are going to explore the OMIT, the OMIA, and switching between slow and fast light in an extended version of the system studied in Ref. \cite{115n}. The proposed system is a hybrid microwave-optomechanical circuit composed of a microwave $LC$ resonator, two SPCTs, and two micromechanical resonators [see Fig. \ref{fig1}(a)]. Under certain conditions the proposed setup is shown to be equivalent to a two-mechanical-modes optomechanical cavity [see Fig. \ref{fig1}(b)] where, in addition to the standard radiation-pressure coupling, the cavity mode interacts with the mechanical modes through the \text{CK}, a higher-order generalized CK, and a three-mode CK type of coupling. Furthermore, there exists an induced CK coupling between the two-mechanical modes. Based on this platform, we focus on how the OMIT, the OMIA, and group delay are affected by the nonlinear \text{CK} couplings in the system. For this purpose, we first consider a simplified version of the proposed configuration in which a single SCPT is coupled to a microwave $LC$ resonator and a single micromechanical resonator. We find that, apart from the optomechanical coupling, the \text{CK} and higher-order nonlinear \text{CK} interactions play a significant role in the emergence and enhancement of transparency window. Then, by considering the original proposed configuration, we find that the presence of the three-mode \text{CK} coupling (which is absent in the simplified configuration) not only gives rise to the appearance of the OMIA effect, but also makes the transparency window wider. Moreover, our findings reveal that one can efficiently control both OMIT and OMIA effects through adjusting the power of the control laser. More interestingly, we show that in both configuration the CK and the generalized \text{CK} couplings can enhance the gain and contribute to the amplification of the output probe field in specific frequency regions. We also examine conditions of slow and fast light behaviors in the proposed system, which are controllable by varying some system parameters. 
\\

The structure of the paper is as follows. In Sec. \ref{sec2}, we introduce the physical system and derive an effective Hamiltonian that describes the radiation-pressure, the phonon-photon CK, the higher-order phonon-photon CK, the three-mode phonon-photon CK, and the phonon-phonon CK couplings. In Sec. \ref{sec 2a}, we provide a discussion of the system dynamics within the framework of the quantum Langevin equations (QLEs), and derive the amplitude of the output probe field by following the standard relation between the input and output fields. We analyze in Sec. \ref{sec3} the induced transparency and induced absorption phenomena in the proposed system. In Sec. \ref{sec4}, we discuss how to control the switch from slow to fast light of the output probe field. Finally, we draw our conclusions in Sec. \ref{sec5}. The details of some derivations are provided in Appendix  \ref{sec6}-\ref{sec9}.  

\section{\label{sec2} SYSTEM MODEL AND EFFECTIVE HAMILTONIAN}

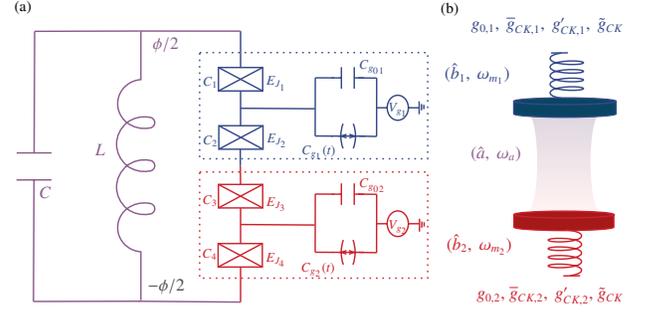
\begin{figure}
	\centering
\begin{adjustbox}{width=0.98\linewidth}  

\tikzset {_643k9gud3/.code = {\pgfsetadditionalshadetransform{ \pgftransformshift{\pgfpoint{0 bp } { 0 bp }  }  \pgftransformrotate{-270 }  \pgftransformscale{2 }  }}}
\pgfdeclarehorizontalshading{_q89dlga25}{150bp}{rgb(0bp)=(1,0.95,0.95);
	rgb(37.5bp)=(1,0.95,0.95);
	rgb(62.5bp)=(0.87,0.84,0.88);
	rgb(100bp)=(0.87,0.84,0.88)}


\tikzset {_x34st4uek/.code = {\pgfsetadditionalshadetransform{ \pgftransformshift{\pgfpoint{-28.5 bp } { 0 bp }  }  \pgftransformrotate{-270 }  \pgftransformscale{2 }  }}}
\pgfdeclarehorizontalshading{_x6sft1i8a}{150bp}{rgb(0bp)=(1,0.95,0.95);
	rgb(37.5bp)=(1,0.95,0.95);
	rgb(62.5bp)=(0.87,0.84,0.88);
	rgb(100bp)=(0.87,0.84,0.88)}
\tikzset{every picture/.style={line width=0.75pt}} 

\begin{tikzpicture}[x=0.75pt,y=0.75pt,yscale=-1,xscale=1]
	
	\draw  [color={rgb, 255:red, 140; green, 90; blue, 142 }  ,draw opacity=1 ] (126.21,275.63) -- (126.58,190.84) (114.48,172.02) -- (138.85,171.98) (114.39,190.86) -- (138.77,190.83) (126.66,172) -- (127.03,87.21) ;
	\draw  [color={rgb, 255:red, 53; green, 73; blue, 135 }  ,draw opacity=1 ] (255.16,112.42) -- (285.98,112.42) -- (285.98,128.8) -- (255.16,128.8) -- cycle ;
	\draw  [color={rgb, 255:red, 53; green, 73; blue, 135 }  ,draw opacity=1 ] (255.41,153.04) -- (286.22,153.04) -- (286.22,169.42) -- (255.41,169.42) -- cycle ;
	\draw  [color={rgb, 255:red, 205; green, 35; blue, 45 }  ,draw opacity=1 ] (255.16,194) -- (285.98,194) -- (285.98,210.38) -- (255.16,210.38) -- cycle ;
	\draw  [color={rgb, 255:red, 205; green, 35; blue, 45 }  ,draw opacity=1 ] (254.92,235.23) -- (285.73,235.23) -- (285.73,251.61) -- (254.92,251.61) -- cycle ;
	\draw [color={rgb, 255:red, 53; green, 73; blue, 135 }  ,draw opacity=1 ]   (255.16,112.42) -- (285.98,128.8) ;
	\draw [color={rgb, 255:red, 205; green, 35; blue, 45 }  ,draw opacity=1 ]   (254.92,235.23) -- (285.73,251.61) ;
	\draw [color={rgb, 255:red, 205; green, 35; blue, 45 }  ,draw opacity=1 ]   (255.16,194) -- (285.98,210.38) ;
	\draw [color={rgb, 255:red, 53; green, 73; blue, 135 }  ,draw opacity=1 ]   (255.41,153.04) -- (286.22,169.42) ;
	\draw [color={rgb, 255:red, 53; green, 73; blue, 135 }  ,draw opacity=1 ]   (255.16,128.8) -- (285.98,112.42) ;
	\draw [color={rgb, 255:red, 205; green, 35; blue, 45 }  ,draw opacity=1 ]   (254.92,251.61) -- (285.73,235.23) ;
	\draw [color={rgb, 255:red, 205; green, 35; blue, 45 }  ,draw opacity=1 ]   (255.16,210.38) -- (285.98,194) ;
	\draw [color={rgb, 255:red, 53; green, 73; blue, 135 }  ,draw opacity=1 ]   (255.41,169.42) -- (286.22,153.04) ;
	\draw [color={rgb, 255:red, 53; green, 73; blue, 135 }  ,draw opacity=1 ]   (270.57,87.08) -- (270.57,112.75) ;
	\draw [color={rgb, 255:red, 53; green, 73; blue, 135 }  ,draw opacity=1 ]   (270.57,128.58) -- (270.57,153.15) ;
	\draw [color={rgb, 255:red, 205; green, 35; blue, 45 }  ,draw opacity=1 ]   (270.37,181.01) -- (270.57,194.11) ;
	\draw [color={rgb, 255:red, 205; green, 35; blue, 45 }  ,draw opacity=1 ]   (270.57,210.49) -- (270.57,235.06) ;
	\draw [color={rgb, 255:red, 205; green, 35; blue, 45 }  ,draw opacity=1 ]   (270.57,251.99) -- (270.57,276.07) ;
	\draw [color={rgb, 255:red, 140; green, 90; blue, 142 }  ,draw opacity=1 ][fill={rgb, 255:red, 19; green, 254; blue, 226 }  ,fill opacity=1 ]   (126.22,275.63) -- (200.25,275.46) ;
	\draw [color={rgb, 255:red, 140; green, 90; blue, 142 }  ,draw opacity=1 ][fill={rgb, 255:red, 19; green, 254; blue, 226 }  ,fill opacity=1 ]   (199.43,275.47) -- (270.57,276.07) ;
	\draw [color={rgb, 255:red, 140; green, 90; blue, 142 }  ,draw opacity=1 ][fill={rgb, 255:red, 19; green, 254; blue, 226 }  ,fill opacity=1 ]   (127.03,87.21) -- (200.35,87) ;
	\draw [color={rgb, 255:red, 140; green, 90; blue, 142 }  ,draw opacity=1 ][fill={rgb, 255:red, 19; green, 254; blue, 226 }  ,fill opacity=1 ]   (200.35,87) -- (270.57,87.08) ;
	\draw [color={rgb, 255:red, 53; green, 73; blue, 135 }  ,draw opacity=1 ]   (270.57,140.87) -- (322.32,141.22) ;
	\draw  [color={rgb, 255:red, 53; green, 73; blue, 135 }  ,draw opacity=1 ] (334.12,119.68) -- (340.84,119.68) (356.52,119.68) -- (349.8,119.68) (340.84,112.18) -- (340.84,127.18) (349.8,112.18) -- (349.8,127.18) ;
	\draw  [color={rgb, 255:red, 53; green, 73; blue, 135 }  ,draw opacity=1 ] (342.02,151.65) .. controls (339.9,156.98) and (339.9,163.55) .. (342.02,168.88) (333.73,160.27) -- (340.36,160.27) (357.4,160.27) -- (350.77,160.27) (349.12,151.65) .. controls (351.23,156.98) and (351.23,163.55) .. (349.12,168.88) ;
	\draw [color={rgb, 255:red, 53; green, 73; blue, 135 }  ,draw opacity=1 ]   (322.65,119.62) -- (322.56,160.21) ;
	\draw [color={rgb, 255:red, 53; green, 73; blue, 135 }  ,draw opacity=1 ]   (365.43,119.8) -- (365.36,160.54) ;
	\draw [color={rgb, 255:red, 53; green, 73; blue, 135 }  ,draw opacity=1 ]   (322.65,119.62) -- (335.47,119.68) ;
	\draw [color={rgb, 255:red, 53; green, 73; blue, 135 }  ,draw opacity=1 ]   (352.4,119.68) -- (365.43,119.8) ;
	\draw [color={rgb, 255:red, 53; green, 73; blue, 135 }  ,draw opacity=1 ]   (322.56,160.21) -- (334.92,160.27) ;
	\draw [color={rgb, 255:red, 53; green, 73; blue, 135 }  ,draw opacity=1 ]   (353.05,160.27) -- (365.36,160.54) ;
	\draw [color={rgb, 255:red, 205; green, 35; blue, 45 }  ,draw opacity=1 ]   (270.57,221.95) -- (322.32,222.31) ;
	\draw  [color={rgb, 255:red, 205; green, 35; blue, 45 }  ,draw opacity=1 ] (334.12,200.77) -- (340.84,200.77) (356.52,200.77) -- (349.8,200.77) (340.84,193.27) -- (340.84,208.27) (349.8,193.27) -- (349.8,208.27) ;
	\draw  [color={rgb, 255:red, 205; green, 35; blue, 45 }  ,draw opacity=1 ] (342.22,232.74) .. controls (340.22,238.07) and (340.22,244.63) .. (342.22,249.96) (334.4,241.35) -- (340.65,241.35) (356.73,241.35) -- (350.48,241.35) (348.92,232.74) .. controls (350.91,238.07) and (350.91,244.63) .. (348.92,249.96) ;
	\draw [color={rgb, 255:red, 205; green, 35; blue, 45 }  ,draw opacity=1 ]   (322.65,200.71) -- (322.56,241.29) ;
	\draw [color={rgb, 255:red, 205; green, 35; blue, 45 }  ,draw opacity=1 ]   (365.43,200.89) -- (365.36,241.62) ;
	\draw [color={rgb, 255:red, 205; green, 35; blue, 45 }  ,draw opacity=1 ]   (322.65,200.71) -- (335.47,200.77) ;
	\draw [color={rgb, 255:red, 205; green, 35; blue, 45 }  ,draw opacity=1 ]   (352.4,200.77) -- (365.43,200.89) ;
	\draw [color={rgb, 255:red, 205; green, 35; blue, 45 }  ,draw opacity=1 ]   (322.56,241.29) -- (334.92,241.35) ;
	\draw [color={rgb, 255:red, 205; green, 35; blue, 45 }  ,draw opacity=1 ]   (353.05,241.35) -- (365.36,241.62) ;
	\draw  [color={rgb, 255:red, 53; green, 73; blue, 135 }  ,draw opacity=1 ] (372.77,140.52) .. controls (372.77,135.7) and (376.07,131.8) .. (380.14,131.8) .. controls (384.21,131.8) and (387.51,135.7) .. (387.51,140.52) .. controls (387.51,145.34) and (384.21,149.24) .. (380.14,149.24) .. controls (376.07,149.24) and (372.77,145.34) .. (372.77,140.52) -- cycle (365.39,140.52) -- (372.77,140.52) (394.89,140.52) -- (387.51,140.52) ;
	\draw [color={rgb, 255:red, 53; green, 73; blue, 135 }  ,draw opacity=1 ]   (395.05,136.01) -- (395.05,145.02) ;
	\draw [color={rgb, 255:red, 53; green, 73; blue, 135 }  ,draw opacity=1 ]   (396.68,142.29) -- (396.68,139.28) -- (396.68,137.92) ;
	\draw [color={rgb, 255:red, 53; green, 73; blue, 135 }  ,draw opacity=1 ]   (398.31,141.47) -- (398.31,138.46) ;
	\draw  [color={rgb, 255:red, 205; green, 35; blue, 45 }  ,draw opacity=1 ] (372.77,221.4) .. controls (372.77,216.46) and (376.07,212.47) .. (380.14,212.47) .. controls (384.21,212.47) and (387.51,216.46) .. (387.51,221.4) .. controls (387.51,226.33) and (384.21,230.33) .. (380.14,230.33) .. controls (376.07,230.33) and (372.77,226.33) .. (372.77,221.4) -- cycle (365.39,221.4) -- (372.77,221.4) (394.89,221.4) -- (387.51,221.4) ;
	\draw [color={rgb, 255:red, 205; green, 35; blue, 45 }  ,draw opacity=1 ]   (395.25,217.91) -- (395.25,226.92) ;
	\draw [color={rgb, 255:red, 205; green, 35; blue, 45 }  ,draw opacity=1 ]   (396.88,224.19) -- (396.88,221.19) -- (396.88,219.82) ;
	\draw [color={rgb, 255:red, 205; green, 35; blue, 45 }  ,draw opacity=1 ]   (398.51,223.37) -- (398.51,220.37) ;
	\draw  [color={rgb, 255:red, 53; green, 73; blue, 135 }  ,draw opacity=1 ][dash pattern={on 0.84pt off 2.51pt}] (242.04,102.04) -- (401,102.04) -- (401,176.09) -- (242.04,176.09) -- cycle ;
	\draw  [color={rgb, 255:red, 205; green, 35; blue, 45 }  ,draw opacity=1 ][dash pattern={on 0.84pt off 2.51pt}] (242.04,185.31) -- (401,185.31) -- (401,259.36) -- (242.04,259.36) -- cycle ;
	\draw [color={rgb, 255:red, 53; green, 73; blue, 135 }  ,draw opacity=1 ]   (341.99,160.27) -- (348.77,160.27) ;
	\draw [shift={(350.77,160.27)}, rotate = 180] [color={rgb, 255:red, 53; green, 73; blue, 135 }  ,draw opacity=1 ][line width=0.75]    (4.37,-1.32) .. controls (2.78,-0.56) and (1.32,-0.12) .. (0,0) .. controls (1.32,0.12) and (2.78,0.56) .. (4.37,1.32)   ;
	\draw [shift={(339.99,160.27)}, rotate = 360] [color={rgb, 255:red, 53; green, 73; blue, 135 }  ,draw opacity=1 ][line width=0.75]    (4.37,-1.32) .. controls (2.78,-0.56) and (1.32,-0.12) .. (0,0) .. controls (1.32,0.12) and (2.78,0.56) .. (4.37,1.32)   ;
	\draw [color={rgb, 255:red, 205; green, 35; blue, 45 }  ,draw opacity=1 ]   (341.99,241.35) -- (348.48,241.35) ;
	\draw [shift={(350.48,241.35)}, rotate = 180] [color={rgb, 255:red, 205; green, 35; blue, 45 }  ,draw opacity=1 ][line width=0.75]    (4.37,-1.32) .. controls (2.78,-0.56) and (1.32,-0.12) .. (0,0) .. controls (1.32,0.12) and (2.78,0.56) .. (4.37,1.32)   ;
	\draw [shift={(339.99,241.35)}, rotate = 0] [color={rgb, 255:red, 205; green, 35; blue, 45 }  ,draw opacity=1 ][line width=0.75]    (4.37,-1.32) .. controls (2.78,-0.56) and (1.32,-0.12) .. (0,0) .. controls (1.32,0.12) and (2.78,0.56) .. (4.37,1.32)   ;
	\draw [color={rgb, 255:red, 53; green, 73; blue, 135 }  ,draw opacity=1 ]   (270.37,169.55) -- (270.37,181.01) ;
	\draw  [color={rgb, 255:red, 140; green, 90; blue, 142 }  ,draw opacity=1 ] (201.08,274.74) -- (201.08,240.96) .. controls (193.93,240.48) and (187.82,235.79) .. (185.68,229.16) .. controls (183.54,222.52) and (185.82,215.29) .. (191.41,210.94) .. controls (195.77,207.59) and (201.41,206.22) .. (206.88,207.19) .. controls (209.02,207.19) and (210.75,208.87) .. (210.75,210.94) .. controls (210.75,213.01) and (209.02,214.7) .. (206.88,214.7) .. controls (201.41,215.66) and (195.77,214.3) .. (191.41,210.94) .. controls (186.76,207.04) and (184.13,201.61) .. (184.13,195.93) .. controls (184.13,190.25) and (186.76,184.82) .. (191.41,180.92) .. controls (195.77,177.57) and (201.41,176.2) .. (206.88,177.17) .. controls (209.02,177.17) and (210.75,178.85) .. (210.75,180.92) .. controls (210.75,182.99) and (209.02,184.67) .. (206.88,184.67) .. controls (201.41,185.64) and (195.77,184.27) .. (191.41,180.92) .. controls (186.76,177.02) and (184.13,171.59) .. (184.13,165.91) .. controls (184.13,160.23) and (186.76,154.8) .. (191.41,150.9) .. controls (195.77,147.55) and (201.41,146.18) .. (206.88,147.15) .. controls (209.02,147.15) and (210.75,148.83) .. (210.75,150.9) .. controls (210.75,152.97) and (209.02,154.65) .. (206.88,154.65) .. controls (201.41,155.62) and (195.77,154.25) .. (191.41,150.9) .. controls (185.82,146.55) and (183.54,139.32) .. (185.68,132.69) .. controls (187.82,126.05) and (193.93,121.36) .. (201.08,120.88) -- (201.08,87.1) ;
	\draw  [draw opacity=0][shading=_q89dlga25,_643k9gud3] (488.14,145.31) -- (520.7,145.31) .. controls (517.28,145.31) and (514.5,160.21) .. (514.5,178.6) .. controls (514.5,196.99) and (517.28,211.89) .. (520.7,211.89) -- (488.14,211.89) .. controls (484.72,211.89) and (481.94,196.99) .. (481.94,178.6) .. controls (481.94,160.21) and (484.72,145.31) .. (488.14,145.31) -- cycle ;
	\draw  [draw opacity=0][shading=_x6sft1i8a,_x34st4uek] (503.67,211.89) -- (471.1,211.9) .. controls (474.53,211.9) and (477.31,196.99) .. (477.3,178.6) .. controls (477.3,160.22) and (474.52,145.31) .. (471.09,145.31) -- (503.66,145.31) .. controls (507.09,145.31) and (509.87,160.21) .. (509.87,178.6) .. controls (509.87,196.98) and (507.1,211.89) .. (503.67,211.89) -- cycle ;
	\draw  [color={rgb, 255:red, 205; green, 32; blue, 45 }  ,draw opacity=1 ][fill={rgb, 255:red, 166; green, 25; blue, 25 }  ,fill opacity=1 ] (530.7,215) -- (530.7,223.77) .. controls (530.7,225.07) and (515.12,226.13) .. (495.9,226.13) .. controls (476.69,226.13) and (461.11,225.07) .. (461.11,223.77) -- (461.11,215)(530.7,215) .. controls (530.7,216.31) and (515.12,217.36) .. (495.9,217.36) .. controls (476.69,217.36) and (461.11,216.31) .. (461.11,215) .. controls (461.11,213.7) and (476.69,212.64) .. (495.9,212.64) .. controls (515.12,212.64) and (530.7,213.7) .. (530.7,215) -- cycle ;
	\draw  [color={rgb, 255:red, 53; green, 73; blue, 135 }  ,draw opacity=1 ] (497.93,134.23) .. controls (492.13,133.86) and (486.33,132.61) .. (486.33,130.11) .. controls (486.33,125.11) and (509.53,125.11) .. (509.53,127.11) .. controls (509.53,129.11) and (486.33,129.11) .. (486.33,124.11) .. controls (486.33,119.11) and (509.53,119.11) .. (509.53,121.11) .. controls (509.53,123.11) and (486.33,123.11) .. (486.33,118.11) .. controls (486.33,113.11) and (509.53,113.11) .. (509.53,115.11) .. controls (509.53,117.11) and (486.33,117.11) .. (486.33,112.11) .. controls (486.33,107.11) and (509.53,107.11) .. (509.53,109.11) .. controls (509.53,111.11) and (486.33,111.11) .. (486.33,106.11) .. controls (486.33,104.16) and (489.84,102.98) .. (494.13,102.37) ;
	\draw  [color={rgb, 255:red, 205; green, 35; blue, 45 }  ,draw opacity=1 ] (496.6,225.7) .. controls (502.4,226.07) and (508.2,227.32) .. (508.2,229.82) .. controls (508.2,234.82) and (485,234.82) .. (485,232.82) .. controls (485,230.82) and (508.2,230.82) .. (508.2,235.82) .. controls (508.2,240.82) and (485,240.82) .. (485,238.82) .. controls (485,236.82) and (508.2,236.82) .. (508.2,241.82) .. controls (508.2,246.82) and (485,246.82) .. (485,244.82) .. controls (485,242.82) and (508.2,242.82) .. (508.2,247.82) .. controls (508.2,252.82) and (485,252.82) .. (485,250.82) .. controls (485,248.82) and (508.2,248.82) .. (508.2,253.82) .. controls (508.2,255.77) and (504.69,256.96) .. (500.4,257.57) ;
	\draw  [color={rgb, 255:red, 53; green, 73; blue, 135 }  ,draw opacity=1 ][fill={rgb, 255:red, 9; green, 69; blue, 99 }  ,fill opacity=1 ] (532.2,135.98) -- (532.2,144.74) .. controls (532.2,146.05) and (516.26,147.1) .. (496.59,147.1) .. controls (476.93,147.1) and (460.99,146.05) .. (460.99,144.74) -- (460.99,135.98)(532.2,135.98) .. controls (532.2,137.28) and (516.26,138.34) .. (496.59,138.34) .. controls (476.93,138.34) and (460.99,137.28) .. (460.99,135.98) .. controls (460.99,134.68) and (476.93,133.62) .. (496.59,133.62) .. controls (516.26,133.62) and (532.2,134.68) .. (532.2,135.98) -- cycle ;
	\draw [color={rgb, 255:red, 205; green, 35; blue, 45 }  ,draw opacity=1 ]   (495.5,257.8) -- (507.2,257.7) ;
	\draw [color={rgb, 255:red, 53; green, 73; blue, 135 }  ,draw opacity=1 ]   (487,102.3) -- (498.7,102.2) ;
	
	\draw (128.58,194.24) node [anchor=north west][inner sep=0.75pt]  [font=\small,color={rgb, 255:red, 140; green, 90; blue, 142 }  ,opacity=1 ]  {$\mathit{C}$};
	\draw (168.04,164.26) node [anchor=north west][inner sep=0.75pt]  [font=\small,color={rgb, 255:red, 140; green, 90; blue, 142 }  ,opacity=1 ]  {$\mathit{L}$};
	\draw (243.05,117.7) node [anchor=north west][inner sep=0.75pt]  [font=\scriptsize,color={rgb, 255:red, 53; green, 73; blue, 135 }  ,opacity=1 ]  {${\textstyle C_{1}}$};
	\draw (242.86,200.15) node [anchor=north west][inner sep=0.75pt]  [font=\scriptsize,color={rgb, 255:red, 205; green, 32; blue, 45 }  ,opacity=1 ]  {${\textstyle C_{3}}$};
	\draw (243.05,158.11) node [anchor=north west][inner sep=0.75pt]  [font=\scriptsize,color={rgb, 255:red, 53; green, 73; blue, 135 }  ,opacity=1 ]  {${\textstyle C_{2}}$};
	\draw (242.64,237.83) node [anchor=north west][inner sep=0.75pt]  [font=\scriptsize,color={rgb, 255:red, 205; green, 32; blue, 45 }  ,opacity=1 ]  {${\textstyle C_{4}}$};
	\draw (287.07,117.15) node [anchor=north west][inner sep=0.75pt]  [font=\scriptsize,color={rgb, 255:red, 53; green, 73; blue, 135 }  ,opacity=1 ]  {${\textstyle E_{J_{1}}}$};
	\draw (286.66,240.01) node [anchor=north west][inner sep=0.75pt]  [font=\scriptsize,color={rgb, 255:red, 205; green, 32; blue, 45 }  ,opacity=1 ]  {${\textstyle E_{J_{4}}}$};
	\draw (287.07,201.24) node [anchor=north west][inner sep=0.75pt]  [font=\scriptsize,color={rgb, 255:red, 205; green, 32; blue, 45 }  ,opacity=1 ]  {${\textstyle E_{J_{3}}}$};
	\draw (287.48,157.01) node [anchor=north west][inner sep=0.75pt]  [font=\scriptsize,color={rgb, 255:red, 53; green, 73; blue, 135 }  ,opacity=1 ]  {${\textstyle E_{J_{2}}}$};
	\draw (351.98,105.47) node [anchor=north west][inner sep=0.75pt]  [font=\scriptsize,color={rgb, 255:red, 53; green, 73; blue, 135 }  ,opacity=1 ]  {${\textstyle C_{g_{0}{}_{1}}}$};
	\draw (352.06,188.42) node [anchor=north west][inner sep=0.75pt]  [font=\scriptsize,color={rgb, 255:red, 205; green, 32; blue, 45 }  ,opacity=1 ]  {${\textstyle C_{g_{0}{}_{2}}}$};
	\draw (311.55,247.26) node [anchor=north west][inner sep=0.75pt]  [font=\scriptsize,color={rgb, 255:red, 205; green, 32; blue, 45 }  ,opacity=1 ]  {${\textstyle C_{g_{2}}( t)}$};
	\draw (311.96,165.33) node [anchor=north west][inner sep=0.75pt]  [font=\scriptsize,color={rgb, 255:red, 53; green, 73; blue, 135 }  ,opacity=1 ]  {${\textstyle C_{g_{1}}( t)}$};
	\draw (372.09,135.81) node [anchor=north west][inner sep=0.75pt]  [font=\scriptsize,color={rgb, 255:red, 53; green, 73; blue, 135 }  ,opacity=1 ]  {${\textstyle V_{g_{1}}}$};
	\draw (372.09,217.26) node [anchor=north west][inner sep=0.75pt]  [font=\scriptsize,color={rgb, 255:red, 205; green, 32; blue, 45 }  ,opacity=1 ]  {${\textstyle V_{g_{2}}}$};
	\draw (207.94,90.04) node [anchor=north west][inner sep=0.75pt]  [font=\small,color={rgb, 255:red, 140; green, 90; blue, 142 }  ,opacity=1 ]  {$\phi /2$};
	\draw (204.49,258.6) node [anchor=north west][inner sep=0.75pt]  [font=\small,color={rgb, 255:red, 74; green, 74; blue, 74 }  ,opacity=1 ]  {$-\phi /2$};
	\draw (111.5,64.5) node [anchor=north west][inner sep=0.75pt]   [align=left] {{\fontfamily{ptm}\selectfont {\footnotesize (a)}}};
	\draw (426.79,75.53) node [anchor=north west][inner sep=0.75pt]  [font=\small,color={rgb, 255:red, 53; green, 73; blue, 135 }  ,opacity=1 ]  {$\ g_{0,1} ,\ \overline{g}_{CK,1} ,\ g_{CK,1}^{\prime } ,\ \tilde{g}_{CK}{}$};
	\draw (411.45,108.06) node [anchor=north west][inner sep=0.75pt]  [font=\small,color={rgb, 255:red, 53; green, 73; blue, 135 }  ,opacity=1 ]  {$(\hat{b}_{1} ,\ \omega _{m_{1}}) \ $};
	\draw (412.73,227.18) node [anchor=north west][inner sep=0.75pt]  [font=\small,color={rgb, 255:red, 205; green, 32; blue, 45 }  ,opacity=1 ]  {$(\hat{b}_{2} ,\ \omega _{m_{2}}) \ $};
	\draw (429.63,165.89) node [anchor=north west][inner sep=0.75pt]  [font=\small,color={rgb, 255:red, 140; green, 90; blue, 142 }  ,opacity=1 ]  {$(\hat{a} ,\ \omega _{a}) \ $};
	\draw (430.82,263.4) node [anchor=north west][inner sep=0.75pt]  [font=\small,color={rgb, 255:red, 53; green, 73; blue, 135 }  ,opacity=1 ]  {$\textcolor[rgb]{0.8,0.13,0.18}{\ g}\textcolor[rgb]{0.8,0.13,0.18}{_{0,2}}\textcolor[rgb]{0.8,0.13,0.18}{,\ }\textcolor[rgb]{0.8,0.13,0.18}{\overline{g}}\textcolor[rgb]{0.8,0.13,0.18}{_{CK,2}}\textcolor[rgb]{0.8,0.13,0.18}{,\ g}\textcolor[rgb]{0.8,0.13,0.18}{_{CK,2}^{\prime }}\textcolor[rgb]{0.8,0.13,0.18}{,\ }\textcolor[rgb]{0.8,0.13,0.18}{\tilde{g}}\textcolor[rgb]{0.8,0.13,0.18}{_{CK}}$};
	\draw (408.67,65.9) node [anchor=north west][inner sep=0.75pt]   [align=left] {{\fontfamily{ptm}\selectfont {\footnotesize (b)}}};	
\end{tikzpicture}

\end{adjustbox}
\caption{(a) Schematic diagram of the proposed microwave-optomechanical system, consisting of two SCPTs with Josephson energies $E_{J_{1(2)}}$ and $E_{J_{3(4)}}$, and respective Josephson capacitances $C_{1(2)}$ and $C_{3(4)}$. These SCPTs are coupled to a common microwave $LC$ resonator and two independent micromechanical resonators characterized by gate capacitances $C_{g_{01}}$ and $C_{g_{02}}$ which couple with time-dependent capacitances $C_{g_{1}}(t)$ and $C_{g_{2}}(t)$. (b) Equivalent cavity optomechanical system representation, where the single cavity mode $\hat{a}$ with frequency $\omega_a$ interacts with two mechanical modes $\hat{b}_1$ and $\hat{b}_2$, with respective natural frequencies  $\omega_{m_1}$ and  $\omega_{m_2}$. The total photon-phonon interaction includes the standard radiation-pressure, the \text{CK}, a higher-order generalized \text{CK}, and a three-mode \text{CK} type of coupling with  coupling strengths $g_{0,1(2)}$, $\bar{g}_{\text{CK},1(2)}$, $g_{\text{CK},1(2)}^\prime$, and  $\tilde{g}_{\text{CK}}$, respectively. Additionally, the two mechanical modes  $\hat{b}_1$ and $\hat{b}_2$ couple with each other via the phonon-phonon \text{CK} interaction with coupling strength $\tilde{g}_{\text{CK}}$. }
	\label{fig1}
\end{figure}

As illustrated in Fig. \ref{fig1}(a), we consider a microwave-optomechanical circuit that consists of two SCPTs characterized by Josephson energies $E_{J_{1(2)}}$ and $E_{J_{3(4)}}$, and corresponding Josephson capacitances $C_{1(2)}$ and $C_{3(4)}$. The two SCPTs are coupled to a common microwave $LC$ resonator and two independent micromechanical resonators with respective gate capacitances $C_{g_{01}}$ and $C_{g_{02}}$ which couple with time-dependent capacitances $C_{g_{1}}(t)$ and $C_{g_{2}}(t)$. A simplified version of the present system, in which a single SCPT is coupled to a microwave $LC$ resonator and a single micromechanical resonator, was previously studied and investigated with the aims of enhancing the radiation-pressure type coupling between the mechanical and microwave modes \cite{36n} and  improving  the few-photon optomechanical effects \cite{115n}. With the same method and under the same approximations used in Ref. \cite{115n}, one can show that the system under consideration can be equivalently described by the Hamiltonian (for details see Appendix \ref{sec6})
\begin{eqnarray}\label{0}
	\hat{H}=\hat{H}_0+\hat{H}_\text{int},
\end{eqnarray}
where
\begin{eqnarray}
	\hat{H}_0&=&\hbar \omega_a \hat{a}^{\dagger}\hat{a}+ \sum_{k=1,2}\hbar\omega_{m_k} \hat{b}_k^{\dagger}\hat{b}_k,\label{01}\\ 
	\hat{H}_\text{int}&=&\sum_{k=1,2}{\hbar}g_{0,k} \hat{a}^{\dagger}\hat{a}\left(\hat{b}_k+\hat{b}_k^{\dagger}\right)+\sum_{k=1,2}{\hbar}\bar{g}_{\text{CK},k}\,\hat{a}^{\dagger}\hat{a}\,\hat{b}_k^{\dagger}\hat{b}_k\label{001}\\
	&&+\sum_{k=1,2}{\hbar}{g}^\prime_{\text{CK},k} \hat{a}^{\dagger}\hat{a}(\hat{b}_k^{\dagger}\hat{b}_k)^2+{\hbar}\tilde{g}_{\text{CK}} \left(2\hat{a}^\dagger \hat{a}+1\right) (\hat{b}_1^\dagger \hat{b}_1) (\hat{b}_2^\dagger \hat{b}_2).	\nonumber
\end{eqnarray}	

The Hamiltonian \ref{0}(a) refers to an (equivalent) optomechanical cavity system where the single cavity mode $\hat{a}$ with frequency $\omega_a $ interacts with two mechanical modes $\hat{b}_1$ and $\hat{b}_2$ with respective effective frequencies $\omega_{m_1}$ and $\omega_{m_2}$ [Fig. \ref{fig1}(b)]. The Hamiltonian of Eq. (\ref{01}) denotes the free energy of the whole system. The interaction Hamiltonian $\hat{H}_\text{int}$ in Eq. (\ref{001}) contains different kinds of nonlinear photon-phonon and phonon-phonon interaction. The cavity mode $\hat{a}$ couples with the mechanical mode $\hat{b}_k\,\, (k=1,2)$ via the radiation-pressure interaction, a higher-order generalized CK interaction (quadratic phonon-number-dependent dispersive shift), a three-mode phonon-photon \text{CK} interaction, and the conventional photon-phonon \text{CK} interaction (linear phonon number-dependent dispersive shift) with respective coupling strengths $g_{0,k}$, $g_{\text{CK},k}^\prime$ , $\tilde{g}_{\text{CK}}$, $\bar{g}_{\text{CK},k}(=g_{\text{CK},k}+g_{\text{CK},k}^\prime+\tilde{g}_{\text{CK}})$. 
Moreover, the two mechanical modes  $\hat{b}_1$ and $\hat{b}_2$ couple with each other via the phonon-phonon \text{CK} interaction with coupling strength $\tilde{g}_{\text{CK}}$. As shown in Appendix \ref{sec6}, the interactions with coupling strengths $g_{0,k}$, $g_{\text{CK},k}$ and $g_{\text{CK},k}^\prime$ originate, respectively, from the quadratic-linear, quadratic-quadratic, and quadratic-quartic coupling terms between the position operator $\hat{x}_c$ of the cavity mode and the position operators  $\hat{x}_{m_k} \,\, (k=1,2)$ of the mechanical modes. In addition, these coupling strengths depend on, among other factors,  $E_J$, $E_c$, $E_J^\prime$, $E_c^\prime$ and $\delta n_{g_{0,1(2)}}$, with $E_c = e^2/[2(C_1 + C_2 + C_{g_{0,1}})]$ and $E_c^\prime = e^2/[2(C_3 + C_4 + C_{g_{0,2}})]$  being the charging energy of each qubit and $\delta n_{g_{0,1(2)}}$ being the deviation from the two lowest charge states integer value (int) $\left|  \text{int} \left(n_{g_{0,k}}\right)\right\rangle =\left|  0_k \right\rangle $ and $\left|  \text{int} \left(n_{g_{0,k}}\right)+1\right\rangle =\left|  1_k \right\rangle \,\,(k=1,2)$, with $n_{g_{0,k}} = V_{g_k}C_{g_{0,k}}/2e$  the gate charge of the $k\text{th}$ Cooper-pair transistor,  by which the energy difference of having zero or one Cooper pairs on the island can be tuned.

We now suppose that the cavity is simultaneously driven by a strong control field with frequency  $\omega_c$ and amplitude $\varepsilon_{c}=\sqrt{\frac{2 \kappa P_{c}}{\hbar \omega_{c}}}$ as well as a weak probe field with frequency $\omega_p$  and amplitude  $\varepsilon_{p}=\sqrt{\frac{2 \kappa P_{p}}{\hbar \omega_{p}}}$ where $P_c$  and  $P_p$ denote the input power of the control and the probe field, respectively, and $\kappa$ is the cavity decay rate. The interaction between the cavity and the driving fields is described by the Hamiltonian
\begin{eqnarray}\label{03}
 	{\hat{H}_\text{dri}}&=&i{\hbar}\left(s_{\text{in}}\hat{a}^{\dagger}-s_{\text{in}}^*\hat{a}\right),
 	\end{eqnarray}
where 	
\begin{eqnarray}\label{030}
	s_{\text{in}}&\equiv&\left(\varepsilon_c+\varepsilon_p e^{-i(\omega_p-\omega_c) t} \right) e^{-i\omega_c t}.
\end{eqnarray}
Therefore, the total Hamiltonian of the considered driven optomechanical system can be written as $	\hat{H}_{total}=\hat{H}_0+\hat{H}_\text{int}+\hat{H}_\text{dri}$. In the rotating frame at the frequency $\omega_c$ of the control field, the system Hamiltonian becomes
\begin{eqnarray}\label{2}
\hat{\tilde{H}}_{total}&=&{\hbar}\Delta_a \hat{a}^{\dagger}\hat{a}+\sum_{k=1,2}{\hbar}\omega_{m_k} \hat{b}_k^{\dagger}\hat{b}_k+\sum_{k=1,2}{\hbar}g_{0,k} \hat{a}^{\dagger}\hat{a}\left(\hat{b}_k+\hat{b}_k^{\dagger}\right)\nonumber\\ 
&&+\sum_{k=1,2}{\hbar}\bar{g}_{\text{CK},k}\,\hat{a}^{\dagger}\hat{a}\,\hat{b}_k^{\dagger}\hat{b}_k+\sum_{k=1,2}{\hbar}{g}^\prime_{\text{CK},k} \hat{a}^{\dagger}\hat{a}(\hat{b}_k^{\dagger}\hat{b}_k)^2\nonumber\\ 
&&+2{\hbar}\tilde{g}_{\text{CK}}\hat{a}^\dagger \hat{a} (\hat{b}_1^\dagger \hat{b}_1) (\hat{b}_2^\dagger \hat{b}_2)+{\hbar}\tilde{g}_{\text{CK}}(\hat{b}_1^\dagger \hat{b}_1) (\hat{b}_2^\dagger \hat{b}_2)\nonumber\\ 
&&+i{\hbar}\left(\left(\varepsilon_c+\varepsilon_p e^{-i(\omega_p-\omega_c) t} \right)\hat{a}^{\dagger}+\text{H.C.}\right),
\end{eqnarray}
where $\Delta_a=\omega_a-\omega_c$ is the detuning between the cavity field and the control field, and H.C. denotes the Hermitian conjugate.  
\section{\label{sec 2a} QUANTUM DYNAMICS AND FLUCTUATIONS}
The dynamics of the system governed by the Hamiltonian in Eq. (\ref{2})  can be described by a set of nonlinear QLEs. Taking into account the mechanical damping as well as the cavity decay, the QLEs for the operators of the cavity and mechanical modes  is obtained as follows:
\begin{eqnarray}
	\dot{\hat{a}} &=& - i\Delta_a \hat{a} - i g_{0,1} \hat{a}\,(\hat{b}_1^\dagger + \hat{b}_1) - i g_{0,2} (\hat{b}_2^\dagger + \hat{b}_2) \label{3} \\ 
	&& - i\bar{g}_{\text{CK},1} \hat{a}\,\hat{b}_1^\dagger\hat{b}_1- i\bar{g}_{\text{CK},2} \hat{a}\,\hat{b}_2^\dagger\hat{b}_2 -i g_{\text{CK},1}^\prime \hat{a}\,(\hat{b}_1^\dagger\hat{b}_1)^2\nonumber\\
	&&-i g_{\text{CK},2}^\prime \hat{a}\,(\hat{b}_2^\dagger\hat{b}_2)^2 -2i\tilde{g}_{\text{CK}}\hat{a} \hat{b}_1^\dagger \hat{b}_1 \hat{b}_2^\dagger \hat{b}_2+\left(\varepsilon_c+\varepsilon_p e^{-i\delta t} \right)\nonumber \\
	&&-\kappa \hat{a}+\sqrt{2\kappa} \hat{a}^{\text{in}},\nonumber \\  
	\dot{\hat{b}}_1
	&=& -i\omega_{m_1}\hat{b}_1 - ig_{0,1}\hat{a}^\dagger\hat{a} - i\bar{g}_{\text{CK},1} \hat{a}^\dagger \hat{a} \hat{b}_1 \label{3i}\\
	&&- 2i g^\prime_{\text{CK},1}\hat{a}^\dagger\hat{a} \hat{b}_1^\dagger\hat{b}_1^2-2i \tilde g_{\text{CK}} \hat{a}^\dagger\hat{a} \hat{b}_1 \hat{b}_2^\dagger\hat{b}_2-i \tilde g_{\text{CK}} \hat{b}_1 \hat{b}_2^\dagger\hat{b}_2\nonumber \\  
	&&- \gamma_1 \hat{b}_1 +\sqrt{2\gamma_1}\hat{b}_1^{\text{in}},\nonumber \\
	\dot{\hat{b}}_2
	&=& -i\omega_{m_2}\hat{b}_2 - ig_{0,2}\hat{a}^\dagger\hat{a} - i\bar{g}_{\text{CK},2} \hat{a}^\dagger \hat{a} \hat{b}_2  \label{3ii}\\
	&&- 2i g^\prime_{\text{CK},2}\hat{a}^\dagger\hat{a} \hat{b}_2^\dagger\hat{b}_2^2-2i \tilde g_{\text{CK}} \hat{a}^\dagger\hat{a} \hat{b}_2 \hat{b}_1^\dagger\hat{b}_1-i \tilde g_{\text{CK}} \hat{b}_2 \hat{b}_1^\dagger\hat{b}_1\nonumber \\
	&&- \gamma_2 \hat{b}_2 +\sqrt{2\gamma_2}\hat{b}_2^{\text{in}}.\nonumber
\end{eqnarray}
Here, $ \gamma_k $ $ (k=1, 2)$ is the damping rate of the $k\,\text{th}$ mechanical mode, and $\delta \equiv \omega_p-\omega_c$. In addition, the vacuum cavity input noise and the Brownian mechanical input noise operators $\hat{o}^{\text{in}}$ $ (o=a, b_1, b_2)$, having zero mean values, satisfy the Markovian correlation functions \cite{116n}:
\begin{subequations}\label{4}
	\begin{align}
	<\hat{a}^{{\text{in}}}(t)\,\hat{a}^{{\text{in}}^\dagger}(t^\prime)>&= \delta(t-t^\prime),\\
	<\hat{b}_k^{\text{in}}(t)\, \hat{b}_k^{{\text{in}}^\dagger}(t^\prime)>&=(\bar{n}_{\text{th},k}+1)\delta(t-t^\prime),  \\
	<\hat{b}_k^{{\text{in}}^\dagger}(t) \, \hat{b}_k^{\text{in}}(t^\prime)>&=\bar{n}_{\text{th},k} \delta(t-t^\prime),
	\end{align}
\end{subequations}
where $\bar{n}_{\text{th},k}=(e^{\frac{\hbar \omega_{m_k}}{k_B T}}-1)^{-1}$ indicates the mean thermal phonon number of the heath bath of the $k\,\text{th}$ mechanical mode at temperature $T$ and $k_B$ is the Boltzmann constant. Since the control field is much stronger than the probe field $\left(\varepsilon_c\gg \varepsilon_p\right)$, we can express each Heisenberg operator in QLEs as the sum of its steady-state mean value and a small fluctuation around the steady-state mean value, i.e., $\hat{o}=o_0+\delta \hat{o}$ ($ o=a,b_k $). In doing so, the mean values at steady state for the cavity field and the two movable mirrors can be obtained from Eqs. (\ref{3})-(\ref{3ii}) by setting all time derivatives to zero. These are found to be
\begin{subequations}\label{6}
	\begin{align}
		a_0 &=  \frac {\varepsilon_c}{\kappa + i\tilde{\Delta}_a}, \label{6a} \\
		b_{k_0} &= \frac {-i\, g_{0,k}\, \vert a_0 \vert ^2}{\gamma_k + i\tilde{\Delta}_{m_k}} \qquad (k=1,2), \label{6b}
	\end{align}
\end{subequations}  
where we have defined
\begin{subequations}\label{70}
	\begin{align}
	\tilde{\Delta}_a&=\Delta_a+g_{0,1} (b_{10}+b_{10}^*)+\bar{g}_{\text{CK},1}\, \vert b_{10} \vert ^2+g^\prime_{\text{CK},1}\, \vert b_{10} \vert^4\nonumber \\
	&\quad+g_{0,2} (b_{20}+b_{20}^*)+\bar{g}_{\text{CK},2}\, \vert b_{2_0} \vert ^2+g^\prime_{\text{CK},2}\, \vert b_{2_0} \vert^4 \nonumber\\
	&\quad+2 \tilde{g}_{\text{CK}}\vert b_{10} \vert ^2\vert b_{20} \vert ^2,\\ 
	\tilde{\Delta}_{m_1}&=\omega_{m_1}+\bar{g}_{\text{CK},1}\, \vert a_0 \vert^2 +2\,g^\prime_{\text{CK},1}\, \vert a_0 \vert^2\,\vert b_{10} \vert^2\nonumber\\
	&\quad+ \tilde{g}_{\text{CK}}\,(2\vert a_0 \vert^2+1)\vert b_{20} \vert^2,\\ 
	\tilde{\Delta}_{m_2}&=\omega_{m_2}+\bar{g}_{\text{CK},2}\, \vert a_0 \vert^2 +2\,g^\prime_{\text{CK},2}\, \vert a_0 \vert^2\,\vert b_{20} \vert^2 \nonumber\\
	&\quad+ \tilde{g}_{\text{CK}}\,(2\vert a_0 \vert^2+1)\,\vert b_{10}\vert^2.
	\end{align}
\end{subequations} 
In addition, by neglecting the smaller terms of second-order fluctuations, the linearized QLEs for the quantum fluctuations are obtained as follows:
\begin{subequations}\label{7}
	\begin{align}
	\delta\dot{\hat{a}}&= -\left(\kappa +i \tilde{\Delta}_a \right) \delta\hat{a}-ig_{\text{eff},1} (\delta \hat{b}_1+\delta \hat{b}_1^\dagger)-ig_{\text{eff},2} (\delta \hat{b}_2+\delta \hat{b}_2^\dagger)\nonumber\\
	&\quad+\varepsilon_p e^{-i\delta t}+\sqrt{2\kappa}\,\hat{a}^{\text{in}},\\   
	\delta\dot{\hat{b}}_1&=-\left(\gamma_1 +i \tilde{\Delta}_{m_1} \right) \delta\hat{b}_1-ig_{\text{eff},1} (\delta \hat{a}+\delta \hat{a}^\dagger)-ig_{11}(\delta \hat{b}_1+\delta \hat{b}_1^\dagger)\nonumber\\
	&\quad-ig_{mm}(\delta \hat{b}_2+\delta \hat{b}_2^\dagger)+\sqrt{2\gamma_1}\,\hat{b}_1^{\text{in}},\\
	\delta\dot{\hat{b}}_2&=-\left(\gamma_2 +i \tilde{\Delta}_{m_2} \right) \delta\hat{b}_2-ig_{\text{eff},2} (\delta \hat{a}+\delta \hat{a}^\dagger)-ig_{22}(\delta \hat{b}_2+\delta \hat{b}_2^\dagger)\nonumber\\ 
	&\quad-ig_{mm}(\delta \hat{b}_1+\delta \hat{b}_1^\dagger)+\sqrt{2\gamma_2}\,\hat{b}_2^{\text{in}},
		\end{align}
\end{subequations}
with
\begin{subequations}\label{74}
	\begin{align}
	g_{\text{eff},1(2)}&= \left(g_{0,1(2)}+\bar{g}_{\text{CK},1(2)}\,\vert b_{10(20)}\vert +2g_{\text{CK},1(2)}^{\prime}\,\vert b_{10(20)}\vert^3\right)\,|a_0| \nonumber\\
	&\quad+\left(2\tilde g_{\text{CK}}\vert b_{20(10)}\vert^2\vert b_{10(20)}\vert \right)\,|a_0|,\\ 
	g_{11(22)}&=2g_{\text{CK},1(2)}^\prime\vert a_0\vert^2\vert b_{10(20)}\vert^2,\\
	g_{mm}&=\tilde{g}_{\text{CK}}(2\vert a_0\vert^2+1)\vert b_{10}\vert \vert b_{20}\vert,\\ 
	g_{0,1}&=-g_{0,2}\equiv  g_0.
	\end{align}
\end{subequations}

By defining the quadrature fluctuation operators  $\delta \hat{X}_o\equiv \frac{\delta \hat{o}+\delta \hat{o}^\dagger}{\sqrt{2}}$ and $\delta \hat{Y}_o\equiv \frac{\delta \hat{o}-\delta \hat{o}^\dagger}{\sqrt{2}i}$ $ (o=a, b_1, b_2) $, the linearized QLEs (\ref{7}) can be concisely expressed in a matrix form as
\begin{eqnarray}\label{8}
	\delta \dot{\hat{u}}(t)=A \delta \hat{u}(t)+\delta \hat{N}(t),
\end{eqnarray}
where $\delta \hat{u}(t)=(\delta \hat{X}_a, \delta \hat{Y}_a, \delta \hat{X}_{b_1}, \delta \hat{Y}_{b_1}, \delta \hat{X}_{b_2}, \delta \hat{Y}_{b_2})^T$  (superscript $T$ denotes matrix transpose) is the vector of continuous-variable fluctuation operators and $\delta \hat{N}(t)=(\sqrt{2\kappa} \hat{X}_a^{\text{in}},\sqrt{2\kappa} \hat{Y}_a^{\text{in}}, \sqrt{2\gamma_1} \hat{X}_{b_1}^{\text{in}},\sqrt{2\gamma_1} \hat{Y}_{b_1}^{\text{in}},\sqrt{2\gamma_2} \hat{X}_{b_2}^{\text{in}},\sqrt{2\gamma_2} \hat{Y}_{b_2}^{\text{in}})$ is the corresponding vector of noises. Here, $\hat{X}_o^{\text{in}} \equiv \frac{\hat{o}^{\text{in}}+\hat{o}^{\text{in} ^\dagger}}{\sqrt{2}}$ and $\hat{Y}_o^{\text{in}} \equiv \frac{\hat{o}^{\text{in}}-\hat{o}^{\text{in}^\dagger}}{\sqrt{2}i} $ $ (o=a, b_1, b_2) $ denote the input noise quadrature operators. Furthermore, the time-independent drift matrix $A$ is given by
\begin{eqnarray}\label{A}
	\mathit{A}= \begin{pmatrix}
		
		-\kappa &\tilde \Delta_a &0 &0 &0 &0 \\
		
		-\tilde \Delta_a &-\kappa &-2g_{\text{eff},1} &0 &-2g_{\text{eff},2} &0\\
		
		0&0& -\gamma_1 &\tilde \Delta_{m_1} &0 &0 \\
		
		-2g_{\text{eff},1} & 0 &-\tilde \Delta_{m_1}-2g_{11} &-\gamma_1 &-2g_{mm} &0 \\
		
		0 &0 &0 &0   & -\gamma_2 & \tilde \Delta_{m_2} \\
		
		-2g_{\text{eff},2} &0 &-2g_{mm} &0 &-\tilde \Delta_{m_2}-2g_{22} & -\gamma_2 \\ 
	\end{pmatrix}.\quad
\end{eqnarray}
On account of the linearized dynamics of the fluctuations and since all noises are Gaussian, the steady state is a zero-mean Gaussian state which is fully characterized by the $6\times 6$ stationary covariance matrix (CM) $V$ with entries
\begin{eqnarray}\label{10}
	V_{ij}=\frac{\left\langle \delta \hat{u}_i(\infty)\delta \hat{u}_j(\infty)+\delta \hat{u}_j(\infty)\delta \hat{u}_i(\infty)\right\rangle}{2}.
\end{eqnarray}
The stability of the steady state of the system can be verified based on the eigenvalues of the CM $V$. To do this, one needs to solve Eq. (\ref{8}) the solution of which  is given by 
\begin{eqnarray}
	\delta \hat{u}(t)=M(t)\,\delta \hat{u}(0)+\int_0^t ds\, M(s)\, \delta \hat{N}(t-s),
\end{eqnarray}
where $M(t)=\text{exp}\left(At\right)$. To ensure the stability of the system, all the eigenvalues of the drift matrix $A$ should have negative real parts. The stability conditions can be derived based on the Routh-Hurwitz criterion \cite{Rou1,Rou2}, imposing constraints on the system parameters. Because of their bulky and tedious expressions, we do not report them here. However, throughout this work, we will check numerically the stability conditions and choose all the parameters in the stable regime. If the stability conditions are satisfied then $M(\infty)=0$, and therefore the steady-state solution is found to be
\begin{eqnarray}
	\delta \hat{u}_i(\infty)=\int_0^\infty ds\, \sum_j M_{ij}(s) \delta\hat{N}_j(t-s),
\end{eqnarray}
by which the steady-state values of the CM elements are obtained as
\begin{eqnarray}\label{m1}
	V_{ij}=\sum_{l,l^\prime}\int_0^\infty dt \int_0^\infty ds M_{il}(t) M_{jl^\prime}(s) D_{ll^\prime}(t-s),
\end{eqnarray}
where $D_{ll^\prime}(t-s)$ are the elements of the diffusion matrix $D(t-s)$ given by
\begin{eqnarray}\label{m2}
	D_{ll^\prime}(t-t^\prime)&=&\frac{\left\langle \delta\hat{N}_i(\infty)\delta\hat{N}_j(\infty)+\delta\hat{N}_j(\infty)\delta\hat{N}_i(\infty)\right\rangle}{2}, \nonumber \\ 
	&&=D_{ll^\prime}\delta(t-t^\prime).
\end{eqnarray}
Substitution Eq. (\ref{m2}) into Eq. (\ref{m1}) results in the stationary state as
\begin{eqnarray}
	V=\int_0^\infty ds\, M(s)\,D\,M^T(s),
\end{eqnarray}
which under the stability conditions leads to the following Lyapunov equation
\begin{eqnarray}\label{111}
	A\,V+V\,A^{T}=-D,
\end{eqnarray}
with
\begin{eqnarray}\label{110}
		D&=&\text{diag}\left(\kappa,\kappa,\gamma_1 n_{\text{th},1},\gamma_1 n_{\text{th},1},\gamma_2 n_{\text{th},2},\gamma_2 n_{\text{th},2}\right),
\end{eqnarray} 
\\
where $n_{\text{th},k}\equiv 2\bar{n}_{\text{th},k}+1 \, (k=1,2)$. In the weak driving regime $\left(\varepsilon_p\ll\varepsilon_c\right)$, the nonlinear interactions between the cavity and the mechanical modes affect the response of the system to the probe field. To investigate the mean response of the present hybrid system, we use a standard procedure in the sideband theory and make the following ansatz for the expectation values  $(\delta a, \delta b_1, \delta b_2)$ of the fluctuations $(\delta \hat{a}, \delta \hat{b}_1, \delta \hat{b}_2)$ (in the rotating frame at the frequency $\omega_c$ of the control field) \cite{31n} 
\begin{subequations}\label{80}
	\begin{align}
	\delta {a}(t)&=A_- e^{-i\delta t}+A_+ e^{i\delta t}, \\ 
	\delta {b}_k(t)&=B_{k-} e^{-i\delta t}+B_{k+} e^{i\delta t}\quad (k=1,2), 
	\end{align}
\end{subequations}
where $O_{\pm} (O=A,B_k)$ represent the amplitudes of the first-order sidebands, with $\pm$ denoting the lower and upper sidebands, respectively.
By inserting these expressions into Eqs.~(\ref{7}) and equating coefficients of terms with the same frequency, we reach the following relations:
\begin{subequations}\label{90}
\begin{align}
	[\kappa+i(\tilde{\Delta}_a-\delta)]A_-&=-ig_{\text{eff},1} (B_{1-}+B_{1-}^*)\nonumber \\
	&\quad-ig_{\text{eff},2} (B_{2-}+B_{2-}^*) +\varepsilon_p,\\ \nonumber
	[\kappa-i(\tilde{\Delta}_a+\delta)]A_-^*&=ig_{\text{eff},1} (B_{1-}+B_{1-}^*)\nonumber \\
	&\quad+ig_{\text{eff},2} (B_{2-}+B_{2-}^*) ,\\ \nonumber
	[\gamma_1+i(\tilde{\Delta}_{m_1}-\delta)]B_{1-}&=-i g_{\text{eff},1} (A_-+A_-^*)-ig_{11}(B_{1-}+B_{1-}^*)\nonumber \\
	&\quad-ig_{mm}(B_{2-}+B_{2-}^*), \\  \nonumber 
	[\gamma_1-i(\tilde{\Delta}_{m_1}+\delta)]B_{1-}^*&=i g_{\text{eff},1} (A_-+A_-^*)+ig_{11}(B_{1-}+B_{1-}^*)\nonumber \\
	&\quad+ig_{mm}(B_{2-}+B_{2-}^*), \\  \nonumber
	[\gamma_2+i(\tilde{\Delta}_{m_2}-\delta)]B_{2-}&=-i g_{\text{eff},2} (A_-+A_-^*)-ig_{22}(B_{2-}+B_{2-}^*)\nonumber \\
	&\quad-ig_{mm}(B_{1-}+B_{1-}^*), \\  \nonumber 
	[\gamma_2-i(\tilde{\Delta}_{m_2}+\delta)]B_{2-}^*&=i g_{\text{eff},2} (A_-+A_-^*)+ig_{22}(B_{2-}+B_{2-}^*)\nonumber \\
	&\quad+ig_{mm}(B_{1-}+B_{1-}^*),
	\end{align}
\end{subequations}
and
\begin{subequations}\label{90s}
\begin{align}
	[\kappa+i(\tilde{\Delta}_a+\delta)]A_+&=-ig_{\text{eff},1} (B_{1+}+B_{1+}^*)\nonumber \\
	&\quad-ig_{\text{eff},2} (B_{2+}+B_{2+}^*) ,\\ \nonumber
	[\kappa-i(\tilde{\Delta}_a-\delta)]A_+^*&=ig_{\text{eff},1} (B_{1+}+B_{1+}^*)\nonumber \\
	&\quad+ig_{\text{eff},2} (B_{2+}+B_{2+}^*)+\varepsilon_p ,\\ \nonumber
	[\gamma_1+i(\tilde{\Delta}_{m_1}+\delta)]B_{1+}&=-i g_{\text{eff},1} (A_++A_+^*)-ig_{11}(B_{1+}+B_{1+}^*)\nonumber \\
	&\quad-ig_{mm}(B_{2+}+B_{2+}^*), \\  \nonumber 
	[\gamma_1-i(\tilde{\Delta}_{m_1}-\delta)]B_{1+}^*&=i g_{\text{eff},1} (A_++A_+^*)+ig_{11}(B_{1+}+B_{1+}^*)\nonumber \\
	&\quad+ig_{mm}(B_{2+}+B_{2+}^*),\\  \nonumber 
	[\gamma_2+i(\tilde{\Delta}_{m_2}+\delta)]B_{2+}&=-ig_{\text{eff},2} (A_++A_+^*)-ig_{22}(B_{2+}+B_{2+}^*)\nonumber \\
	&\quad-ig_{mm}(B_{1+}+B_{1+}^*), \\  \nonumber 
	[\gamma_2-i(\tilde{\Delta}_{m_2}-\delta)]B_{2+}^*&=ig_{\text{eff},2} (A_++A_+^*)+ig_{22}(B_{2+}+B_{2+}^*)\nonumber \\
	&\quad+ig_{mm}(B_{1+}+B_{1+}^*).
	\end{align}
\end{subequations}
The solution for the amplitude $A_-$ is obtained as (we will not list the solutions for other amplitudes one by one, because here we just care about the field with frequency $\omega_p$ in the output field)
\begin{eqnarray}\label{44}
	A_-&=&\frac{1+i\,f(\delta)\, }{\kappa-i(\delta-\tilde{\Delta}_a)-2\tilde{\Delta}_a\,f(\delta)}\varepsilon_p,
\end{eqnarray}
where
\begin{eqnarray}\label{45}
	f(\delta)=-\left(\frac{\xi_1\,g_{\text{eff},1}+\xi_2\,g_{\text{eff},2}}{\xi_0\left(\kappa-i(\delta+\tilde\Delta_a)\right)}\right),
	\end{eqnarray}
with 
\begin{subequations}\label{46}
\begin{align}
	\xi_0&=\left((\gamma_1-i\delta)^2+\tilde{\Delta}_{m_1}(\tilde{\Delta}_{m_1}+2g_{11})\right)\\ 
	 &\quad\times \left((\gamma_2-i\delta)^2+\tilde{\Delta}_{m_2}(\tilde{\Delta}_{m_2}+2g_{22})\right)-4\tilde{\Delta}_{m_1}\tilde{\Delta}_{m_2}\,g_{mm}^2,\nonumber\\ 
    \xi_1&=-2\tilde{\Delta}_{m_1}g_{\text{eff},1}\left((\gamma_2-i\delta)^2+\tilde{\Delta}_{m_2}(\tilde{\Delta}_{m_2}+2g_{22})\right)\\ 
    &\quad+4\tilde{\Delta}_{m_1}\tilde{\Delta}_{m_2}g_{\text{eff},2}\,g_{mm},\nonumber\\ 
    \xi_2&=-2\tilde{\Delta}_{m_2}g_{\text{eff},2}\left((\gamma_1-i\delta)^2+\tilde{\Delta}_{m_1}(\tilde{\Delta}_{m_1}+2g_{11})\right)\\ 
    &\quad+4\tilde{\Delta}_{m_2}\tilde{\Delta}_{m_1}g_{\text{eff},1}\,g_{mm}.\nonumber
	\end{align}
\end{subequations}\label{47}

To calculate the amplitude of the cavity output field, we use the well-known input-output relation \, $\hat{a}^{\text{out}}+\hat{a}^{\text{in}}=\sqrt{2 \kappa}\, \hat{a}$ \cite{in-out} by which we obtain 
\begin{eqnarray}\label{11}
	\varepsilon_{out}+\varepsilon_c e^{-i\omega_c t}+\varepsilon_p e^{-i\omega_p t}=2 \kappa \,(a_0+\delta {a}) e^{-i\omega_c t}.
\end{eqnarray}
By substituting Eq. (\ref{80}a) into Eq. (\ref{11}), it is clear that $A_-$ and $A_+$ oscillate at frequencies $\omega_p$ and $2\omega_c-\omega_p$, respectively. The total output field $\varepsilon_t$ oscillating at the probe frequency $\omega_p$ is given by
\begin{eqnarray}\label{12}
	\varepsilon_t=\frac{2 \kappa \vert A_-\vert }{\varepsilon_p}.
\end{eqnarray}
The dispersive and absorptive properties of the output field at the probe frequency are characterized by the imaginary part 
$\varepsilon_i=\text{Im}\left[\varepsilon_t\right]$ and the real part  $\varepsilon_r=\text{Re}\left[\varepsilon_t\right]$
of the field amplitude $\varepsilon_t$, respectively, \cite{4} which can be measured via, e.g., homodyne detection \cite{in-out}. 

\section{ Optomechanically induced transparency and Absorption}\label{sec3}

In this section, we analyze the induced transparency and absorption phenomena in the proposed optomechanical circuit. We first consider a simplified version of the system where a single SCPT is coupled to a microwave $LC$ resonator and a single micromechanical resonator. In this case, the equivalent optomechanical system is a single-mechanical-mode optomechanical cavity in which one of the end mirrors is movable. We demonstrate that, in addition to the optomechanical interaction responsible for the emergence of OMIT, both the \text{CK} coupling and the generalized \text{CK} coupling play roles in enhancing the width of the transparency window.
Subsequently, we explore the scenario involving two SCPTs and two micromechanical resonators, which is equivalent to an optomechanical system where a single cavity mode is coupled to two movable mirrors, resulting in OMIA. Furthermore, we demonstrate that the three-mode phonon-photon \text{CK} coupling between the cavity and mechanical modes plays an important role in the broadening of the transparency windows that occur in the OMIA profile. 
\subsection{ Single movable mirror case  }\label{sec3a}
\begin{figure}
	\includegraphics[width=0.38\textwidth]{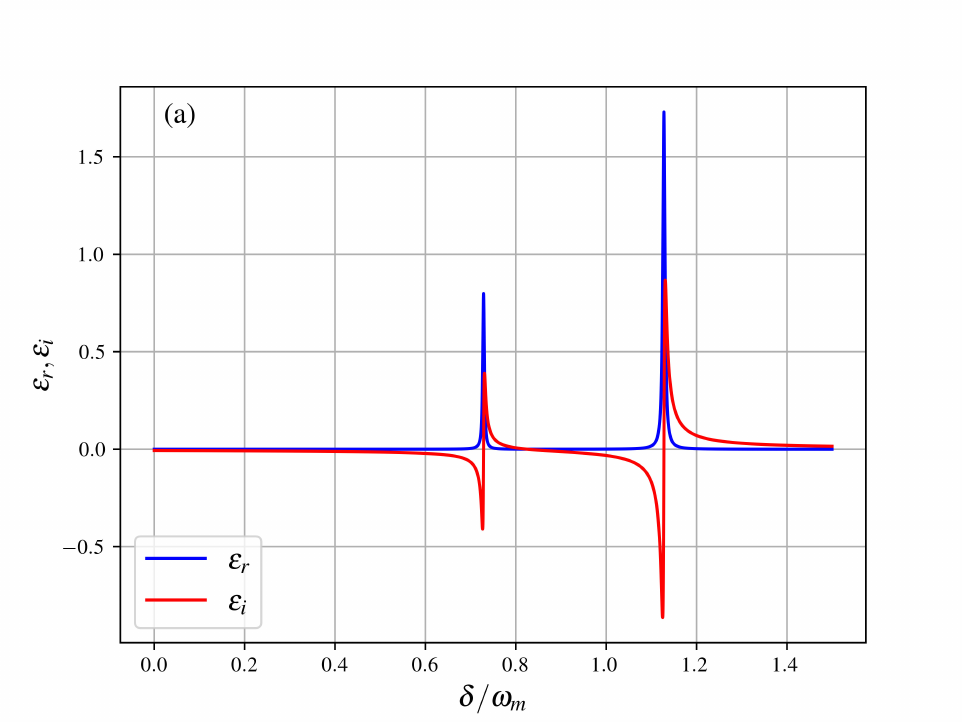}
	\hspace{8mm}
	\includegraphics[width=0.38\textwidth]{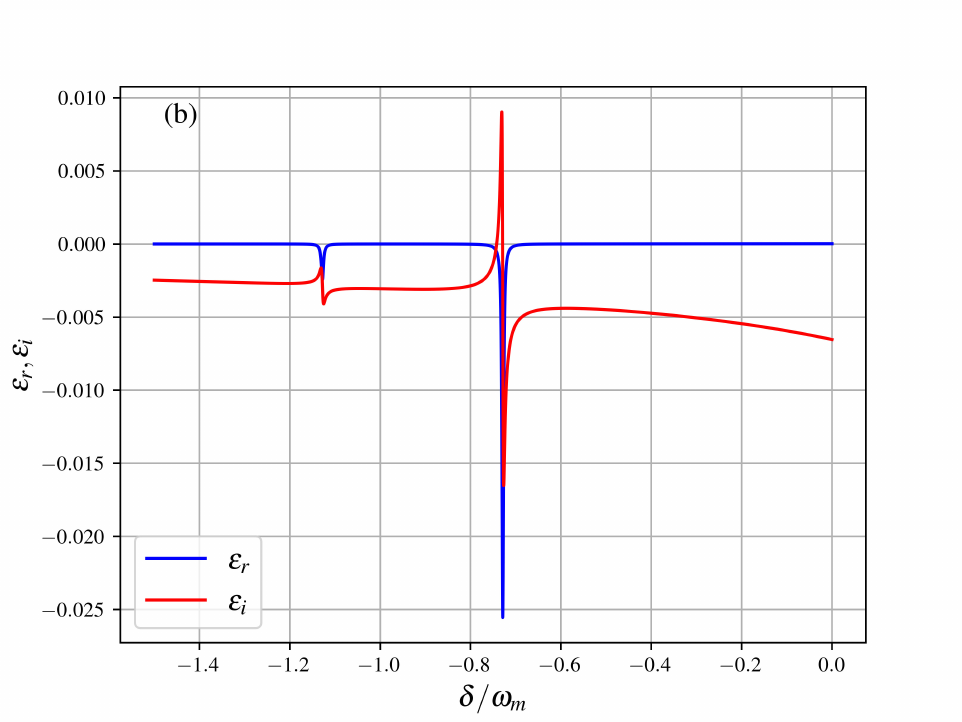}
    \hspace{8mm}
	\includegraphics[width=0.38\textwidth]{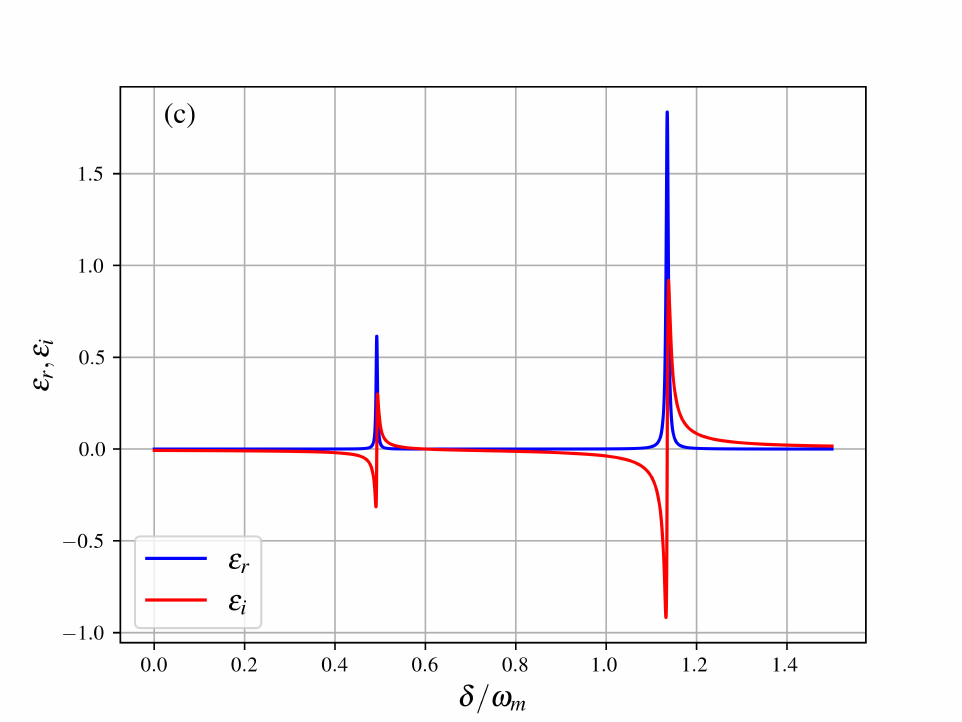}
	\hspace{8mm}
	\includegraphics[width=0.38\textwidth]{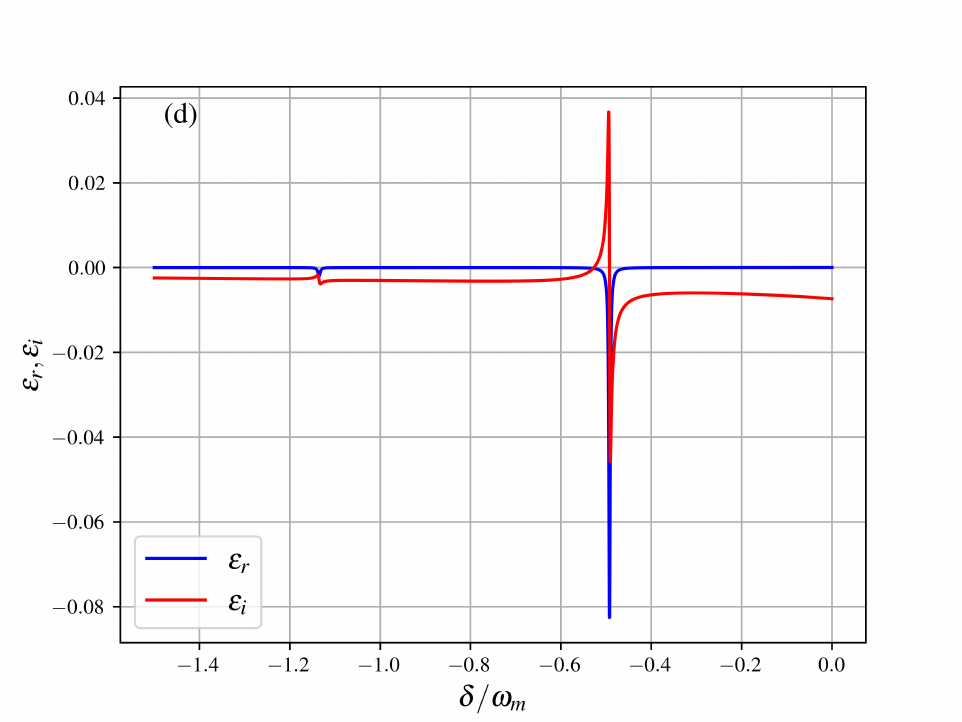}
	\caption{Plots of the absorption (blue lines) and dispersion (red lines) of the probe field against the normalized detuning ${\delta}/{\omega_{m}}$  for two values of the input power of the control field: [(a) and (b)] $P_c=0.07\,\rm{nW}$, [(c) and (d)] $P_c=0.09\, \rm{nW}$. Panels (a) and (c) correspond to the case $\tilde \Delta_a\approx \tilde \Omega_m$ while panels (b) and (d) correspond to the case $\tilde \Delta_a\approx -\tilde \Omega_m$.  Here, the equivalent optomechanical system is a single-mechanical-mode optomechanical cavity in which one of the end mirrors is movable. The experimental values of the parameters are $\omega_c/2 \pi= 10\, \rm{GHz}$, $\omega_m/2 \pi=  50 \, \rm{MHz}$, $\kappa=1\, \rm{MHz}$, $\gamma=500 \, \rm{kHz}$, and $\Delta_a=\omega_{m_1}$ \cite{100n}.} 
	\label{fig10}
\end{figure}
\begin{figure}
	\includegraphics[width=0.4\textwidth]{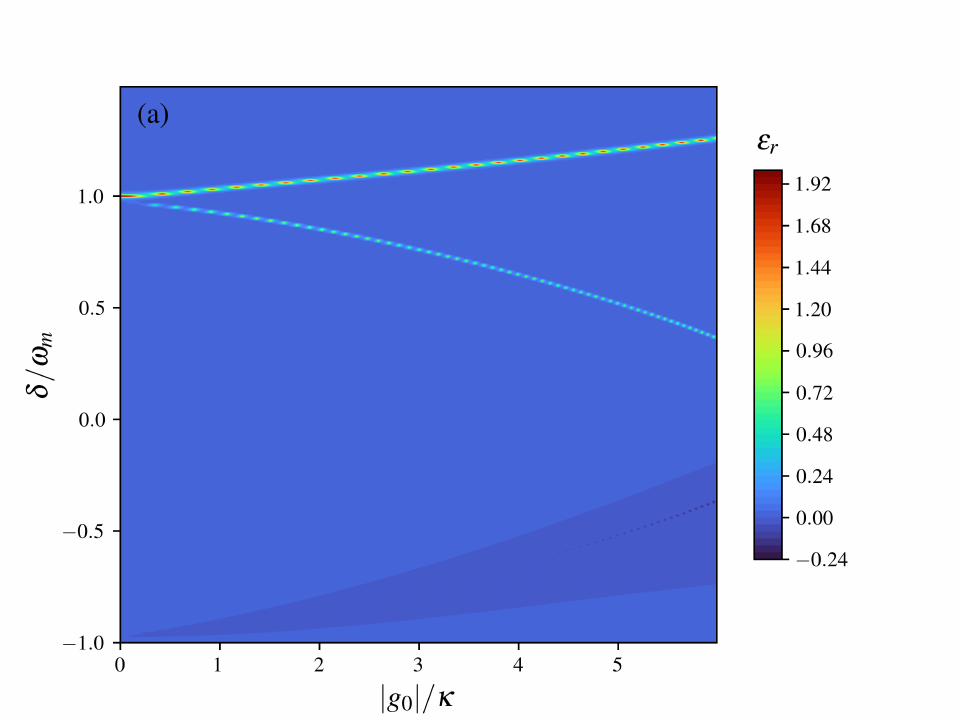}
	\hspace{8mm}
	\includegraphics[width=0.4\textwidth]{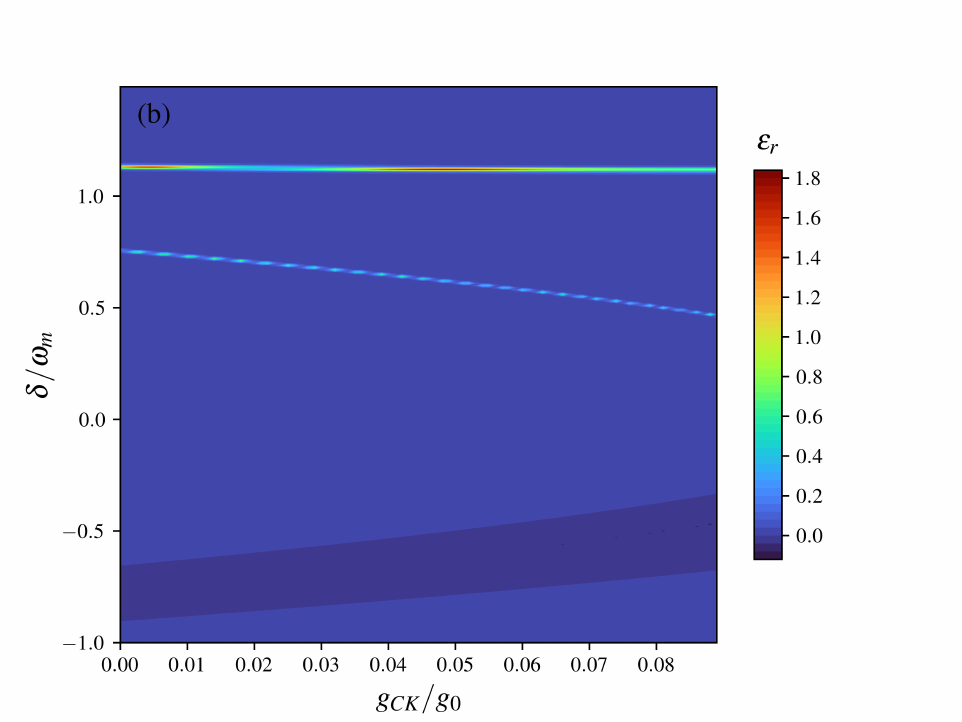}
	\hspace{8mm}
	\includegraphics[width=0.4\textwidth]{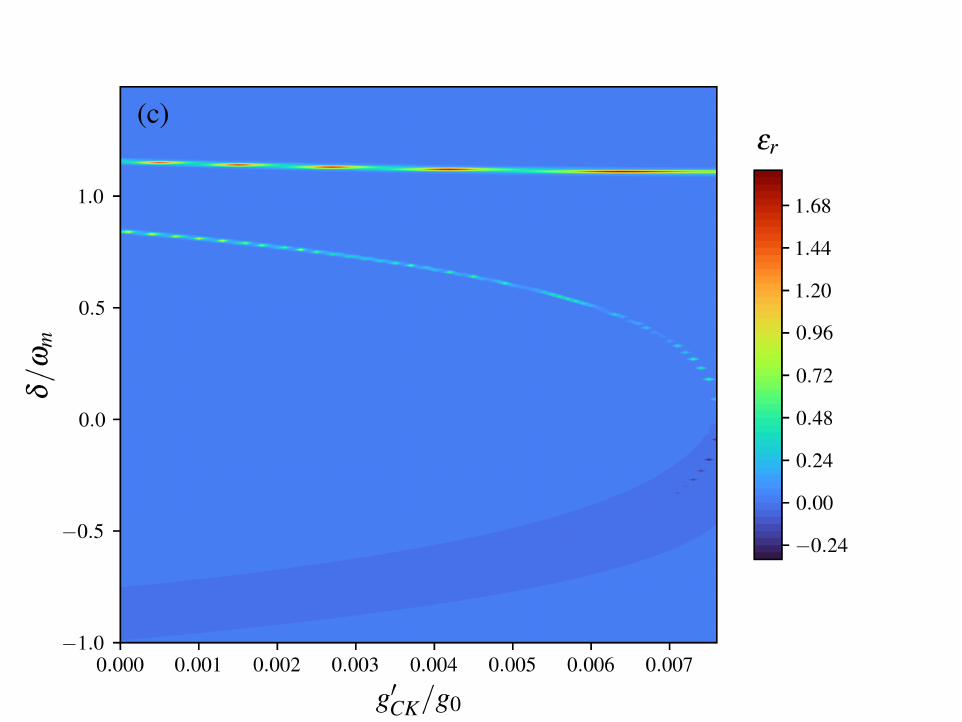}
	\caption{Density plots of the real part $(\varepsilon_r)$ of the field amplitude $\varepsilon_t$ versus the normalized frequency $\delta/\omega_m$ and (a) optomechanical coupling $\vert g_{0} \vert/\kappa$, (b) normalized \text{CK} coupling $g_{\text{CK}}/g_0$, and (c) normalized generalized \text{CK} coupling $g_{\text{CK}}^\prime/g_0$. Here, the equivalent optomechanical system is the same as that in Fig. \ref{fig10}. The parameters are selected the same as Fig. \ref{fig10} with $P_c\sim -70 \, \rm{dBm}\, (0.07\, \rm{nW})$ 
	\cite{100n}.}
	\label{fig2}
\end{figure}
\begin{figure}
	\includegraphics[width=0.45\textwidth]{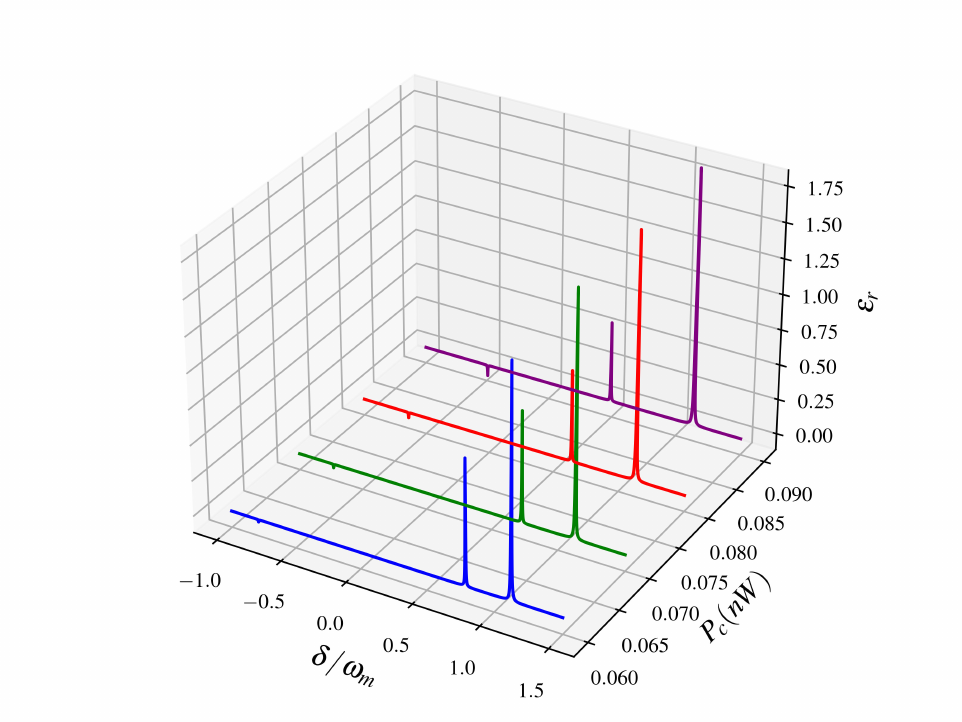}
	\caption{The real part $(\varepsilon_r)$ of the field amplitude $\varepsilon_r$ versus the normalized detuning $\delta/\omega_m$ and the control laser power $P_c$. Here, the equivalent optomechanical system is the same as that in Fig. \ref{fig10}. Other parameters are the same as those in Fig. \ref{fig2}.}
	\label{fig3}
\end{figure}
Considering the case in which a single SCPT is coupled to a microwave $LC$ resonator and a single micromechanical resonator, and assuming that the amplitudes and frequencies of the driving fields (coupling and probe fields) are different, the response of the output field $\varepsilon_t$ is simplified to that of a cavity with one movable mirror,
\begin{eqnarray}\label{121}
	\varepsilon_t= \frac{2 \kappa\,(1+i\,f(\delta)) }{\kappa-i(\delta-\tilde{\Delta}_a)-2\tilde{\Delta}_a\,f(\delta)},
\end{eqnarray}
with
\begin{subequations}\label{122}
	\begin{align}
	f(\delta)&=\frac{2\tilde \Delta_m g_{\text{eff}}^2}{\left((\gamma-i\delta)^2+\tilde \Omega_m^2\right)\left(\kappa-i(\delta+\tilde\Delta_a)\right)},\\ \tilde{\Delta}_a&=\Delta_a+g_{0} (b_{0}+b_{0}^*)+\bar{g}_{\text{CK}}\, \vert b_{0} \vert ^2+g^\prime_{\text{CK}}\, \vert b_{0} \vert^4, \\
\tilde \Omega_m^2 &\equiv \tilde \Delta_m^2+2g_{11}\tilde\Delta_m,\\
\tilde{\Delta}_{m}&=\omega_{m}+\bar{g}_{\text{CK}}\, \vert a_0 \vert^2 +2\,g^\prime_{\text{CK}}\, \vert a_0 \vert^2\,\vert b_{0} \vert^2.
	\end{align}
\end{subequations} 
where $\omega_m=\omega_{m_1}$, $\gamma=\gamma_1$, $b_0=b_{10}$, $g_0=g_{0,1}$ and $\bar g_{\text{CK}}=g_{\text{CK}}+g_{\text{CK}}^\prime$ ( $g_{\text{CK}}=g_{\text{CK},1}$ and $g_{\text{CK}}^\prime=g_{\text{CK},1}^\prime$). 

It can be obviously seen from Eq. (\ref{121}) that the absorption and  dispersion properties of the probe field depend on several system parameters in a complicated nonlinear way. Here, we consider the two cases of $\tilde \Delta_a\approx\pm \tilde \Omega_m$  for which the coupling between the moving mirror and the cavity field becomes stronger. The width $\Gamma_{\text{OMIT}}$ (full width at half maximum) of the transparency window is identified as an important index in OMIT. Following the derivation detailed in Appendix \ref{sec8}, we obtain the following approximate analytical expression for the width  in the resolved sideband regime and for $\tilde \Delta_a\approx \tilde \Omega_m$ \begin{eqnarray}\label{11c}
	\Gamma_{\text{OMIT}}&=&\gamma+\frac{g_{\text{eff}}^2}{\kappa}=\gamma\left(1+C_{\text{eff}}\right),
\end{eqnarray}
where $	g_{\text{eff}}= \left(g_{0}+\bar{g}_{\text{CK}}\,|b_{0}| +2g_{\text{CK}}^\prime|b_{0}|^3\right)\,|a_0|$. The width of the transparency window can thus be expressed as the sum of the intrinsic mechanical damping rate $\gamma$ and the effective damping rate $\gamma_{\text{eff}}=\gamma C_{\text{eff}}$ where $C_{\text{eff}}$ can be seen as an effective collective optomechanical cooperativity parameter defined by $C_{\text{eff}}=\frac{g_{\text{eff}}^2}{\gamma\kappa}$. The dependence of $g_{\text{eff}}$ on the nonlinear coupling strengths $g_{\text{CK}}^\prime$ and $\bar{g}_{\text{CK}}$  provides an additional flexibility for adjusting the width of the transparency window. We point out that in the absence of the \text{CK} and the generalized \text{CK} couplings( i.e., $\bar{g}_{\text{CK}}=g_{\text{CK}}^\prime=0$) Eq. (\ref{11c}) is reduced to the familiar result for the width of the transparency window in a standard optomechanical cavity with the cooperativity parameter $C=\frac{g_0^2}{\gamma \kappa}|a_0|^2$  \cite{31n,77}. Obviously, when the \text{CK} and the generalized \text{CK} couplings are switched on (i.e., $\bar{g}_{\text{CK}}\neq 0$ and  $g_{\text{CK}}^\prime\neq0$) the width of the transparency windows grows considerably wider due to $C_{\text{eff}}\gg C$. On the other hand, the maximum value of the slope of the dispersion curve at the transparency window is given by (see Appendix \ref{sec8}) 	
\begin{eqnarray}\label{12c}
	\text{K}_{\text{max}}=-\frac{2g_{\text{eff}}^2/\kappa}{\left(\gamma+g_{\text{eff}}^2/\kappa\right)^2}=-\frac{1}{\gamma}\frac{2C_{\text{eff}}}{\left(1+C_{\text{eff}}\right)^2}.
\end{eqnarray}
We observe that the transparency window is accompanied by a negative dispersion slope. In Sec. \ref{sec4}, we will show that such negative dispersion behavior can lead to the slow light in the system. In addition, from Eqs. (\ref{11c}) and (\ref{12c}) we get
	\begin{eqnarray}\label{13c}
	\Gamma_{\text{OMIT}}\times \text{K}_{\text{max}}&=&\frac{-2C_{\text{eff}}}{1+ C_{\text{eff}}}\approx-2,
\end{eqnarray}
since $C_{\text{eff}}\gg1$. This shows that the product of the width of the transparency window and the maximum value of the slope of the dispersion curve at the transparency window is a constant; the narrower the width $\Gamma_{\text{OMIT}}$ is, the steeper the dispersion curve becomes. On the other hand, for the case $\tilde \Delta_a\approx -\tilde \Omega_m$, following the lines as above one can similarly obtain the width of the transparency window and the maximum value of the slope of the dispersion curve, respectively as 
\begin{eqnarray}\label{14c}
	\Gamma_{\text{OMIT}}&=&\frac{g_{\text{eff}}^2}{\kappa}-\gamma=\gamma\left(C_{\text{eff}}-1\right),\nonumber\\
	\text{K}_{\text{max}}&=&\frac{1}{\gamma}\frac{2C_{\text{eff}}}{\left(C_{\text{eff}}-1\right)^2},
\end{eqnarray}
from which we can deduce that
\begin{eqnarray}\label{130c}
	\Gamma_{\text{OMIT}}\times \text{K}_{\text{max}}&=&\frac{2C_{\text{eff}}}{ C_{\text{eff}}-1}\approx2.
\end{eqnarray}
As will be shown in Sec \ref{sec4}, the positive dispersion slope causes the fast light effect in the system. 

To demonstrate the absorption and dispersion properties of the probe field in the present case, we have made some simulations using the following experimental parameter setting \cite{100n}  $\omega_c/2 \pi= 10\, \rm{GHz}$, $\omega_m/2 \pi=  50 \, \rm{MHz}$, $\Delta_a=\omega_{m}$, $\kappa=1\, \rm{MHz}$, $\gamma=500 \, \rm{kHz}$, and $P_c\sim -70 \, \rm{dBm}$. In Fig.  \ref{fig10} we have plotted the absorption (blue curves) and dispersion (red curves) of the probe field against the normalized detuning $\delta/\omega_m$ for two values of the input power of the control field, $P_c=0.07\,\rm{nW}$ [panels (a) and (b)] and  $P_c=0.09\,\rm{nW}$ [panels (c) and (d)]. Figures \ref{fig10}(a) and \ref{fig10}(c), which correspond to the case $\tilde\Delta_a\approx\tilde \Omega_m$ show that the probe absorption peak of the output spectrum splits into two peaks, due to the \text{CK} and the generalized \text{CK} couplings, leading to a relatively wide transparency window accompanied by a negative dispersion slope. As one can see by comparing Figs. \ref{fig10}(a) and \ref{fig10}(c), with increasing the control field power the width of the transparency window increases, while the (negative) slope of the dispersion curve decreases. This is consistent with the result of Eq. (\ref{13c}). On the other hand, Figs. \ref{fig10}(b) and \ref{fig10}(d), which correspond to the case $\tilde\Delta_a\approx-\tilde \Omega_m$ show that the situation changes significantly such that the \text{CK} and the generalized \text{CK} couplings give rise to gain profiles $\left(\varepsilon_r<0\right)$ instead of the absorption features. This effect is due to the energy transfer from the control field to the probe field, which is significantly influenced by the presence of nonlinear interactions such as the \text{CK} and generalized \text{CK} couplings. These nonlinear interactions can enhance the gain and contribute to the amplification of the output probe field in specific frequency regions. Moreover, the slope of the dispersion is seen to be positive in the region between the gain peaks, indicating fast light propagation. Also, we see that with increasing the power of the control field the width of the transparency window increases, while the (positive) slope of the dispersion curve decreases, as predicted by Eq. (\ref{130c}).

Now, we are going to examine the impacts of the coupling strengths, including the optomechanical, the  \text{CK}, and the generalized \text{CK} couplings on the absorption of the output probe field. 
\\

In Fig. \ref{fig2}, we plot the real part $(\varepsilon_r)$ of the field amplitude $\varepsilon_t$ as a function of the normalized detuning $\delta/\omega_m$ and (a) normalized optomechanical coupling $\vert g_{0} \vert/\kappa$, (b) normalized \text{CK} coupling $g_{\text{CK}}/g_0$, and (c) normalized generalized \text{CK} coupling $g_{\text{CK}}^\prime/g_0$. We consider the experimentally feasible system parameters as in Fig. \ref{fig10}. Additionally, we have set  $g_{\text{CK}}=-3.3\times0.01\, \rm{MHz} $, $g_{\text{CK}}^\prime=-3.3\times0.003\, \rm{MHz} $ in panel (a); $g_0=-3.3 \, \rm{MHz}$, $g_{\text{CK}}^\prime=0.003\,  g_0 $ in panel (b); and $g_0=-3.3 \, \rm{MHz}$, $g_{\text{CK}}=0.01\, g_0 $ in panel (c). In Fig. \ref{fig2}(a), we see that by turning on the optomechanical interaction, a transparency window (i.e., $\varepsilon_r \simeq 0$) appears whose width increases with the optomechanical coupling strength.
Figures. \ref{fig2}(b) and \ref{fig2}(c) illustrate the impacts of the \text{CK} and the generalized \text{CK} couplings on the probe field absorption profile, respectively. Although both nonlinear interactions lead to the increasing of the width of the transparency window, the generalized \text{CK} coupling can enhance this width more than the \text{CK} coupling due to its presence in two terms (the fourth and fifth terms) of Hamiltonian (\ref{2}). It is also worth noting that for negative values of the normalized detuning $\delta/\omega_m$, the absorption profile shows negative values, indicating the occurrence of amplification. 
\\
\begin{figure}
	\includegraphics[width=0.38\textwidth]{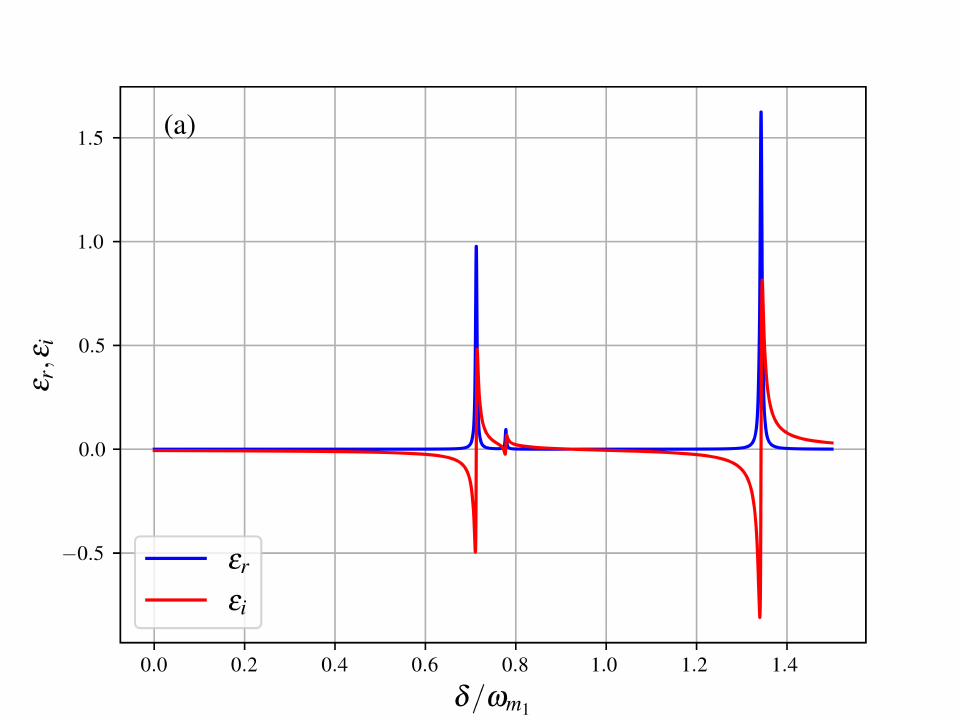}
	\hspace{8mm}
	\includegraphics[width=0.38\textwidth]{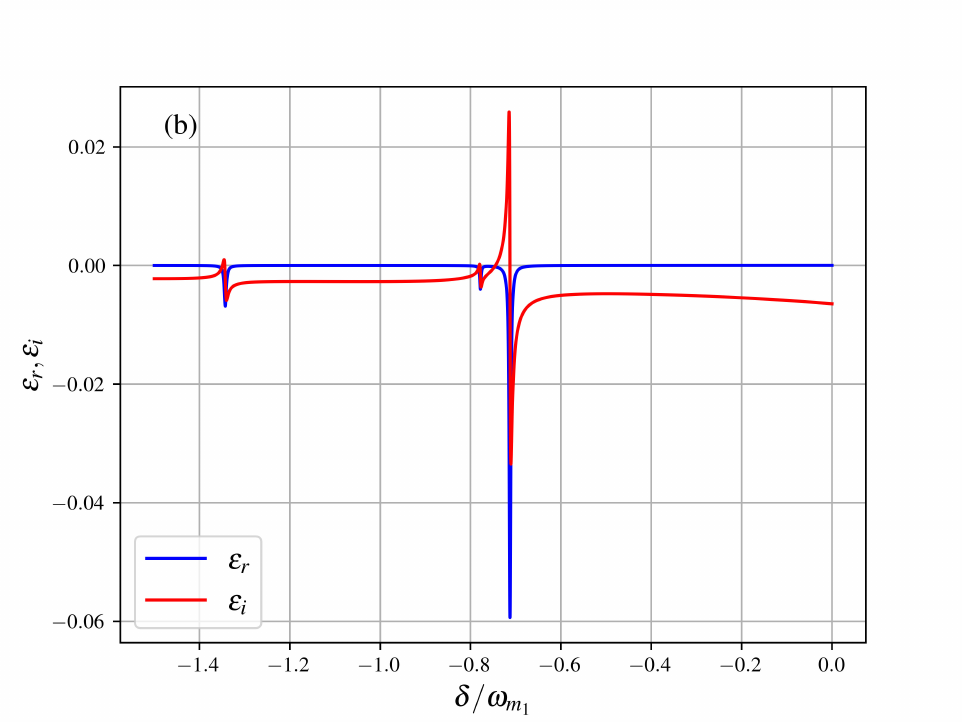}
	\hspace{8mm}
	\includegraphics[width=0.38\textwidth]{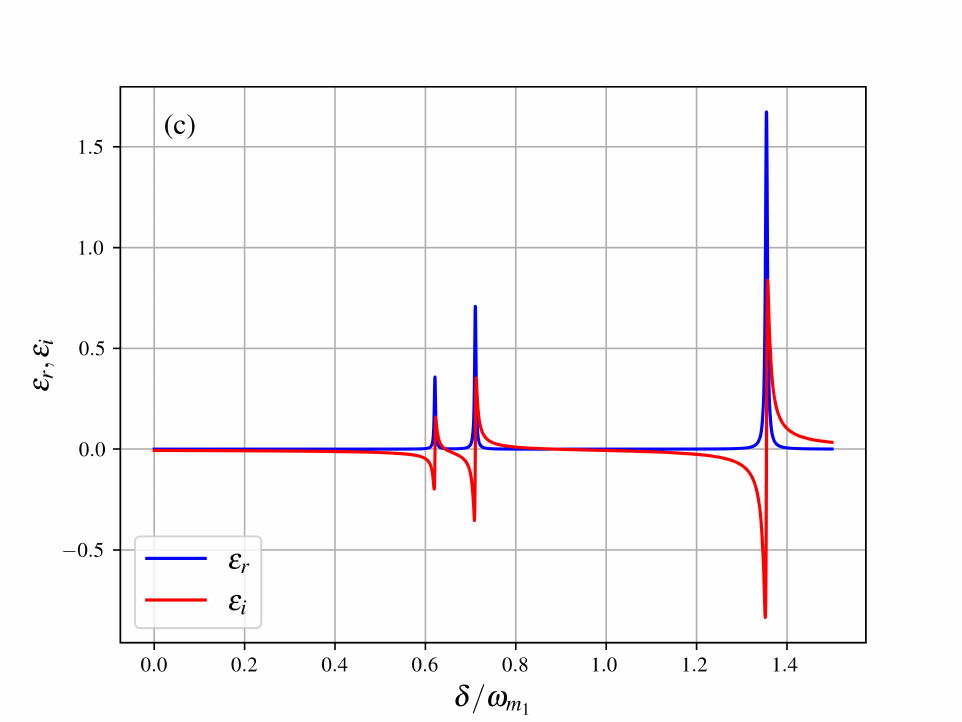}
	\hspace{8mm}
	\includegraphics[width=0.38\textwidth]{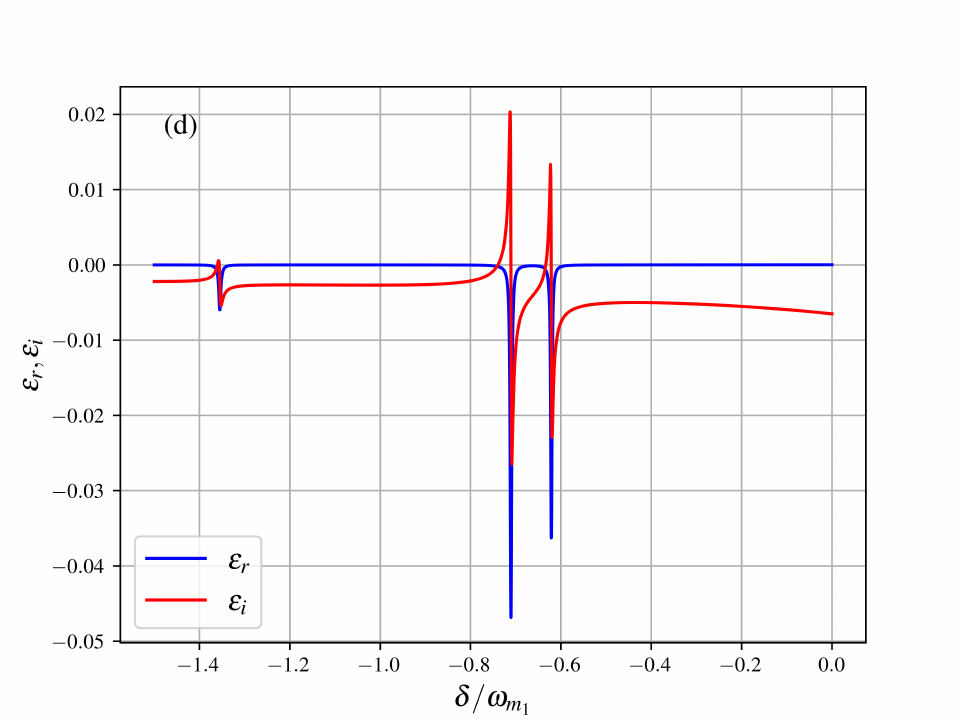}
	\caption{Plots of the absorption (blue lines) and dispersion (red lines) of the probe field against the normalized detuning ${\delta}/{\omega_{m_1}}$  for two values of the input power of the control field: [(a) and (b)] $P_c=0.08\,\rm{nW}$, [(c) and (d)] $P_c=0.09\, \rm{nW}$. Panels (a) and (c) correspond to the case $\tilde \Delta_a\approx \tilde \Omega_{m_k}$ while panels (b) and (d) correspond to the case $\tilde \Delta_a\approx -\tilde \Omega_{m_k}$.  Here, the equivalent optomechanical system is a two-mechanical-modes optomechanical cavity in which
	both end mirrors are movable mirrors, as shown in Fig. \ref{fig1}(b). The experimental values of the parameters are $\omega_c/2 \pi= 10\, \rm{GHz}$, $\omega_{m_1}/2 \pi=\omega_{m_2}/2 \pi=  50 \, \rm{MHz}$, $\kappa=1\, \rm{MHz}$, $\gamma=500 \, \rm{kHz}$, and $\Delta_a=\omega_{m_1}$ \cite{100n}.} 
	\label{fig20}
\end{figure}

To examine the impact of the control field on the probe field absorption profile, we plot in  Fig. \ref{fig3}  the quantity $\varepsilon_r$ versus the normalized detuning $\delta/\omega_m$ and the laser power $P_c$. This figure clearly shows that with the increase of $P_c$, the width of transparency window increases.  Physically, by increasing the control laser power, the amplitudes of the photon and phonon modes increase leading to an increase in the effective coupling strengths. 
\begin{figure}
	\includegraphics[width=0.4\textwidth]{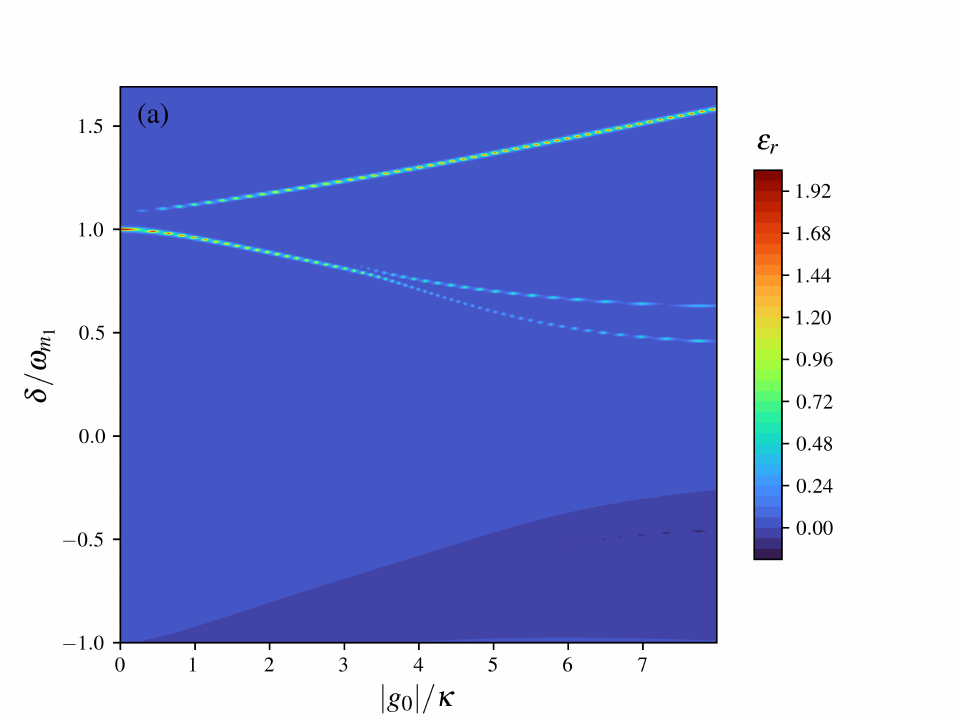}
		\hspace{8mm}
	\includegraphics[width=0.4\textwidth]{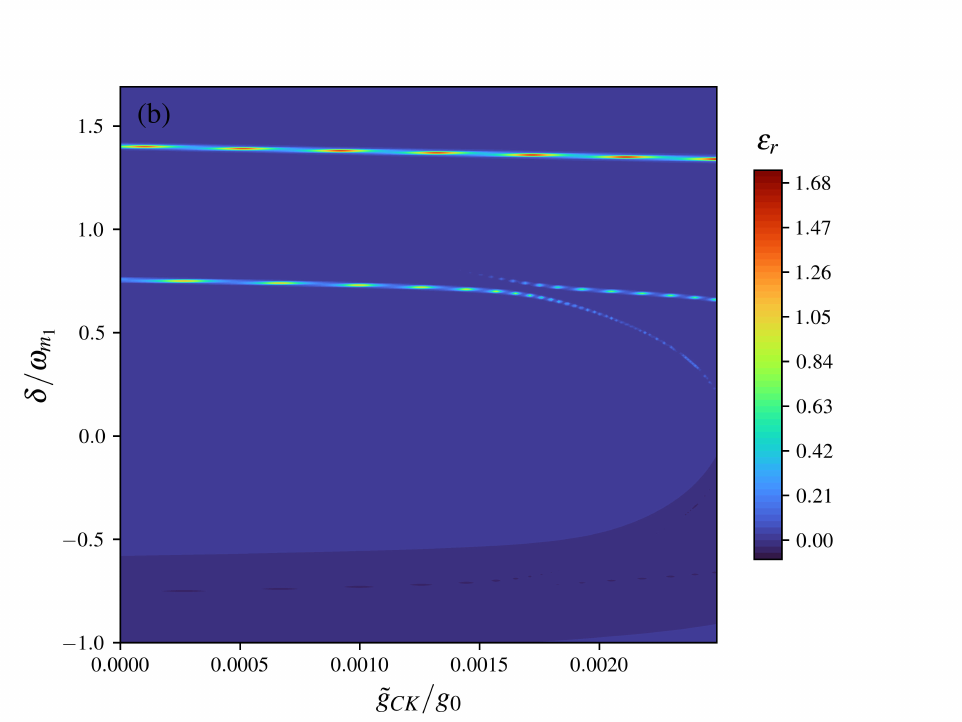}
	\caption{ Density plots of the absorption profile, $\varepsilon_r$, versus the normalized detuning $\delta/\omega_{m_1}$ and (a) optomechanical coupling $\vert g_{0}\vert/\kappa$, (b) normalized three-mode \text{CK} coupling $\tilde g_{\text{CK}}/g_{0}$. Here, the equivalent optomechanical system is the same as that in Fig. \ref{fig20}. The parameters are selected the same as Fig. \ref{fig20} with $P_c\sim -70 \rm{dBm}\, (0.09 \rm{nW})$.} 
		\label{fig4}
\end{figure}
\begin{figure}
	\includegraphics[width=0.45\textwidth]{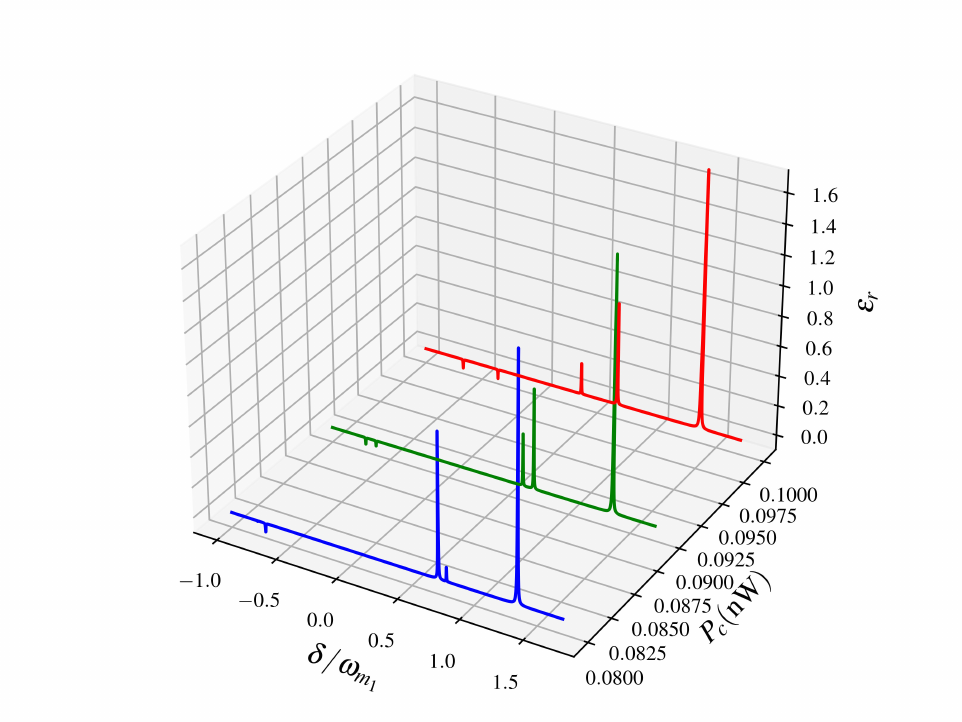}
	\caption{The absorption profile, $\varepsilon_r$, versus the normalized detuning $\delta/\omega_m$  and the control laser power $P_c$. Here, the equivalent optomechanical system is the same as that in Fig. \ref{fig20}. Other parameters are the same as those in Fig. \ref{fig4}.}
	\label{fig5}
\end{figure}
\begin{figure}
	\includegraphics[width=0.45\textwidth]{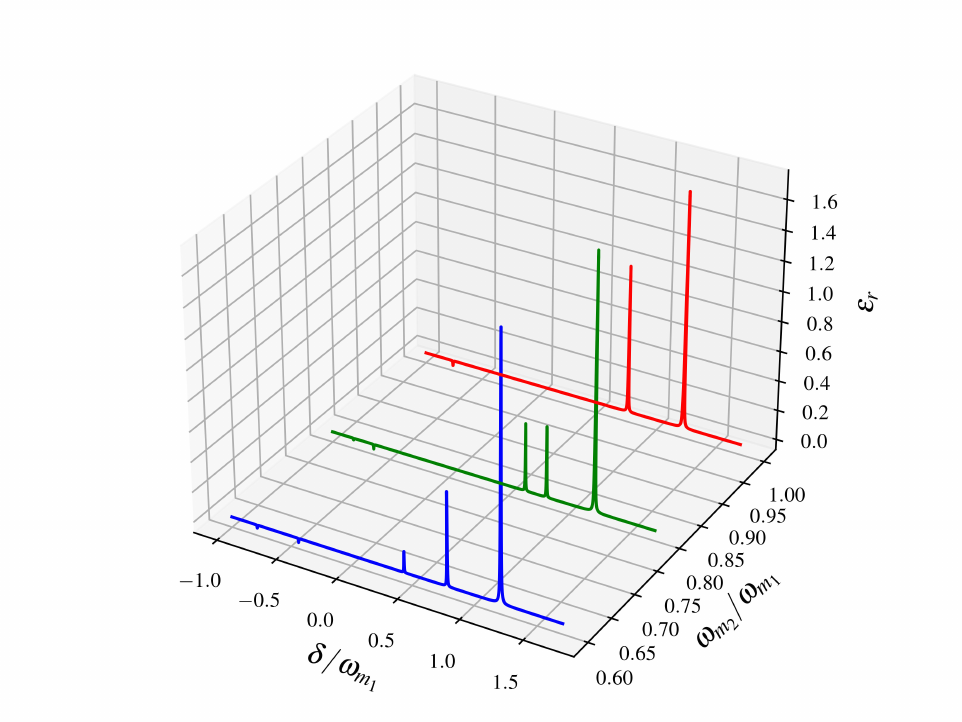}
	\caption{The absorption profile, $\varepsilon_r$, versus the normalized detuning $\delta/\omega_{m_1}$  and the ratio $\omega_{m_2}/\omega_{m_1}$. Here, the equivalent optomechanical system is the same as that in Fig. \ref{fig20}. Other parameters are the same as those in Fig. \ref{fig4}.}
	\label{fig5w}
\end{figure}
\subsection{ Two movable mirrors case }\label{sec3b}
Now, we consider the case in which a common microwave $LC$ resonator two SCPTs are coupled to two independent micromechanical resonators. The new aspect
of this situation is the occurrence of tripartite coupling between the cavity mode and the two mechanical modes through the three-mode \text{CK} coupling (the term proportional to $\hat{a}^\dagger \hat{a} \hat{b}_1^\dagger \hat{b}_1\hat{b}_2^\dagger \hat{b}_2$ in Hamiltonian (\ref{2})), as well as the \text{CK} coupling between the two mechanical modes (the term proportional to $\hat{b}_1^\dagger \hat{b}_1\hat{b}_2^\dagger \hat{b}_2$ in Hamiltonian (\ref{2})) both with coupling strength $\tilde{g}_{\text{CK}}$. Here, we suppose that the two mechanical modes are the same. As can be seen from relations \ref{15app}(a)-(h) in Appendix \ref{sec6}, in this case, the optomechanical, the \text{CK} and the higher-order nonlinear \text{CK} coupling strengths are, respectively, $g_{0,1}=-g_{0,2}\equiv g_{0}$, $g_{\text{CK},1}=g_{\text{CK},2}\equiv g_{\text{CK}}$, $g_{\text{CK},1}^\prime=g_{\text{CK},2}^\prime \equiv g_{\text{CK}}^\prime$, and $\tilde g_{\text{CK}}=2g_{\text{CK}}^\prime$. 

As in the case of the single-mode-mechanical optomechanical system, discussed in part 
A of this section, we can obtain the width of the transparency window and the slope of dispersion curve for the present case. Choosing $\tilde \Delta_a\approx \tilde \Omega_{m_k}$ the results are obtained as follows (see Appendix \ref{sec9})
\begin{eqnarray}\label{8a}
	\Gamma_{\text{OMIT}}&=&\gamma+\frac{g_{\text{eff,1}}^2+g_{\text{eff,2}}^2}{\kappa}\equiv\gamma\left(1+C_{\text{eff}}\right),\label{8aa}\\
	\text{K}_{\text{max}}&=&\frac{-2\left(g_{\text{eff,1}}^2+g_{\text{eff,2}}^2\right)/\kappa}{\left(\gamma+\left(g_{\text{eff,1}}^2+g_{\text{eff,2}}^2\right)/\kappa\right)^2}\equiv-\frac{1}{\gamma}\frac{2C_{\text{eff}}}{\left(1+C_{\text{eff}}\right)^2}.\quad\label{8aaa}
\end{eqnarray}
where $C_{\text{eff}} \equiv \frac{g_{\text{eff,1}}^2+g_{\text{eff,2}}^2}{\gamma \kappa}$  and $g_{\text{eff},i}\, (i=1,2)$ are given in Eq. (\ref{74}a). On the other hand, if we choose $\tilde \Delta_a\approx -\tilde \Omega_{m_k}$, we get
\begin{eqnarray}\label{140}
	\Gamma_{\text{OMIT}}&=&\gamma\left(C_{\text{eff}}-1\right),\label{140a}\\
	\text{K}_{\text{max}}&=&\frac{1}{\gamma}\frac{2C_{\text{eff}}}{\left(C_{\text{eff}}-1\right)^2}.\label{140aa}
\end{eqnarray}
These results are similar to those we obtained for the single-optomechanical-mode case [Eqs. (\ref{11c}), (\ref{12c}) and (\ref{14c})] except for a difference in effective cooperativities. This difference is due to the presence of two additional couplings, i.e., the three-mode photon-phonon CK and the phonon-phonon CK couplings, in the case of two-mechanical modes-optomechanical cavity. Accordingly, we expect that these additional couplings affect the absorption and dispersion profiles of the output probe field.
\\ 

In Fig. \ref{fig20} we have plotted the absorption (blue curves) and dispersion (red curves) of the probe field against the normalized detuning $\delta/\omega_{m_{1}}$ for two values of the input power of the control field, $P_c=0.08\,\rm{nW}$ [panels (a) and (b)] and $P_c=0.09\,\rm{nW}$ [panels (c) and (d)]. By comparing Figs. \ref{fig10} and \ref{fig20}, we can see that in the presence of two mechanical modes the system exhibits two transparency windows in the probe absorption spectrum. As is shown in Fig. \ref{fig20}(a) and \ref{fig20}(c), which correspond to the case $\tilde \Delta_a\approx \tilde \Omega_{m_k}$ increasing the control field power broadens both transparency windows, while the (negative) slopes of the dispersion curves associated with each window decrease.  This behavior is consistent with the expected results from Eqs. (\ref{8aa}) and (\ref{8aaa}). On the other hand, similar to the previous configuration, Figs. \ref{fig20}(b) and \ref{fig20}(d), which correspond to the case $\tilde\Delta_a\approx-\tilde \Omega_{m_k}$ illustrate that the \text{CK} nonlinear interactions can induce gain features $\left(\varepsilon_r<0\right)$, since these nonlinear interactions enhance energy transfer from the control field to the probe field, leading to amplification in specific frequency regions. The positive slope of the dispersion between the gain peaks indicates regions of fast light propagation, similar to what is observed in the single movable mirror case of Fig. \ref{fig10}, but now occurring for both transparency windows. In addition, as before, we see that by increasing the power of the control field the widths of the transparency windows increase, while the (positive) slopes of the dispersion curves decrease.
\\
To illustrate the impacts of the coupling strengths on the absorption of the output probe field, we plot $\varepsilon_r$ versus the normalized detuning $\delta/\omega_{m_1}$ and the normalized optomechanical coupling $\vert g_{0}\vert/\kappa$ in Fig. \ref{fig4}(a) and the normalized three-mode \text{CK} coupling $\tilde g_{\text{CK}}/g_0$ in Fig. \ref{fig4}(b) for $\omega_c/2\pi= 10\, \rm{GHz}$, $\omega_{m_1}/2\pi=\omega_{m_2}/2\pi= 50\, \rm{MHz}$, $\Delta_a=\omega_{m_{1}}$, $\kappa=1\, \rm{MHz}$, $\gamma_1=\gamma_2=500\,\rm{kHz}$, $P_c\sim -70 \rm{dBm} (0.09 \rm{nW})$  \cite{115n}, 
$ g_0=-4.8\, \rm{MHz}$, $g_{\text{CK}}=-0.02\,g_{0} $, $g_{\text{CK}}^\prime=0.0009\,  g_{0} $, and $\tilde g_{\text{CK}} = 2 g_{\text{CK}}^\prime$. 
Interestingly, as can be seen, the additional mechanical mode leads to the appearance of OMIA and we have two transparency windows. Furthermore, we observe that the absorption profile exhibits amplification in a wide range of negative values for $\delta/\omega_{m_1}$. To gain further insight, we show the role of the three-mode \text{CK} coupling in occurrence of the OMIA in the introduced system. As demonstrated in Fig.  \ref{fig4}(b), with increasing this coupling, the widths of the transparency windows increase. Additionally, in Fig. \ref{fig5}, the absorption profile is plotted versus the normalized detuning $\delta/\omega_{m_1}$ and the control laser power $P_c$. This figure clearly shows that by enhancing the control laser power, the widths of transparency windows increase. Therefore, by adjusting the controllable system parameters (as discussed in Appendix \ref{sec6}), one can manipulate and control the transparency windows.

To explore the controllability of the OMIA in the system under consideration, in Fig. \ref{fig5w} the absorption profile, $\varepsilon_r$, is plotted versus the normalized detuning $\delta/\omega_{m_1}$  and the ratio $\omega_{m_2}/\omega_{m_1}$,  $ g_{0,1}=-3.1 \rm{MHz} $, $g_{\text{CK},1}\simeq0.02\,g_{0,1} $, $g_{\text{CK},1}^\prime=0.0003\,  g_{0,1} $. Clearly, by selecting different frequencies for mechanical modes (blue and  green lines), two transparency windows appear. For these parameters, if both mechanical mode frequencies are equal, only OMIT is observed (red line).    
\section{ TUNABLE SLOW AND FAST LIGHT CONVERSION }\label{sec4}
\begin{figure}
	\includegraphics[width=0.4\textwidth]{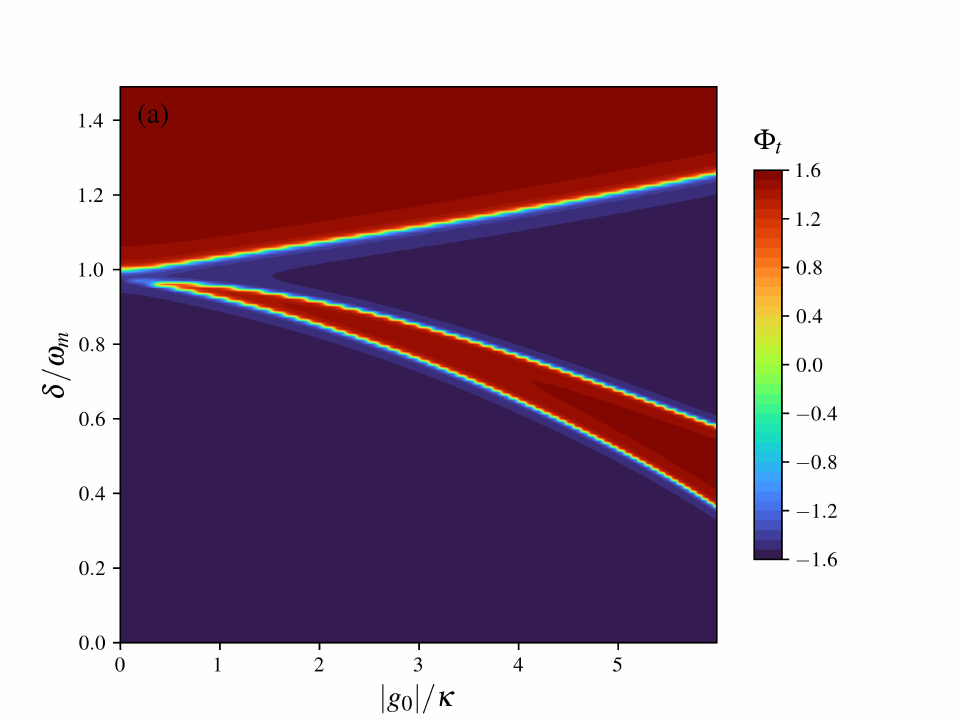}
	\hspace{8mm}
	\includegraphics[width=0.4\textwidth]{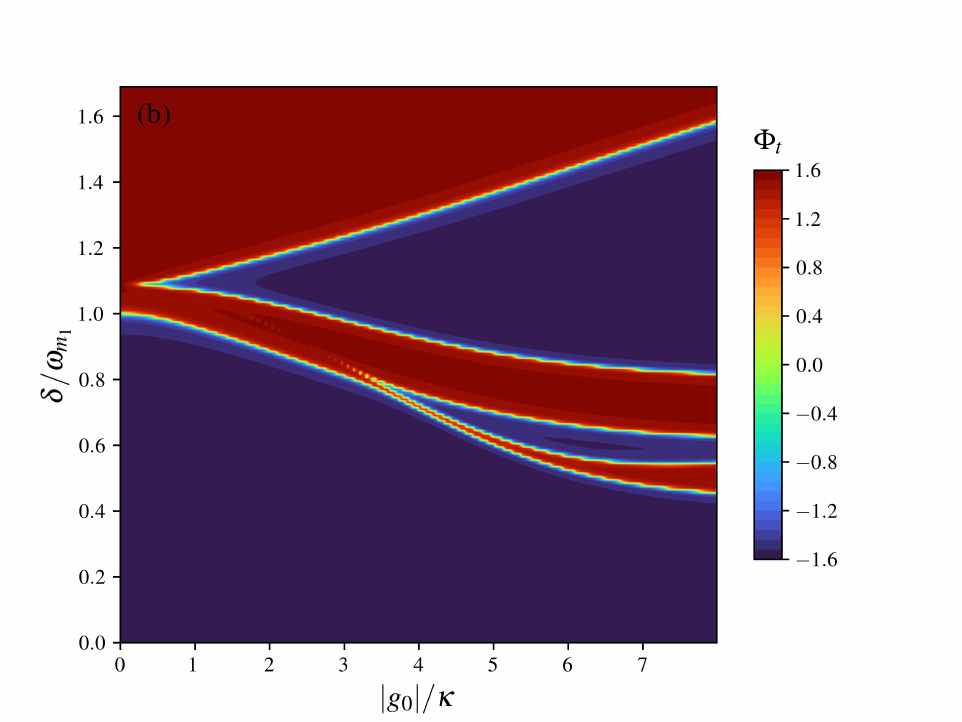}
	\caption{The probe field phase, $ \Phi_t(\omega_p) $, versus the  optomechanical coupling  $\vert g_{0} \vert/\kappa$ and (a) normalized frequency $\delta/\omega_m$ (single-mechanical-mode configuration) and (b) $\delta/\omega_{m_1}$ (two-mechanical-modes configuration). The parameters in panels (a) and (b) are the same as those in Figs. \ref{fig2} and \ref{fig4},
	respectively.}
	\label{fig6}
\end{figure}
In the previous section, we described the OMIT and OMIA, and analyzed how the optomechanical, the \text{CK}, the generalized \text{CK}, and three-mode \text{CK} couplings affect the widths of transparency windows. As we know, OMIT and OMIA enable precise control over light propagation. Thus, in this section, we focus on analyzing the group delay of the total output field at the probe frequency. For this purpose, we use the rapid phase dispersion $\Phi_t(\omega_p)=arg\left[\varepsilon_t(\omega_p)\right]$, and accordingly, the group delay that is expressed as \cite{25n,118}:
\begin{eqnarray}
	\tau_g=\frac{\partial \Phi_t(\omega_p)}{\partial \omega_p}=\text{Im} \left[\frac{1}{\varepsilon_t}\frac{\partial \varepsilon_t}{\partial \omega_p} \right].
\end{eqnarray}
\begin{figure}
	\includegraphics[width=0.45\textwidth]{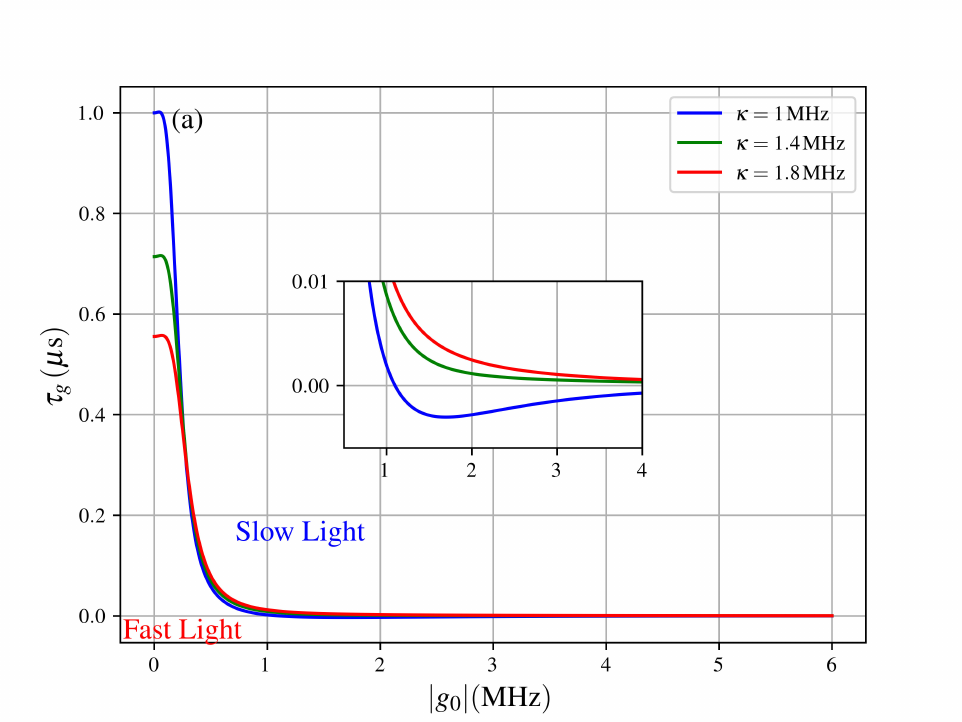}
	\hspace{8mm}
	\includegraphics[width=0.45\textwidth]{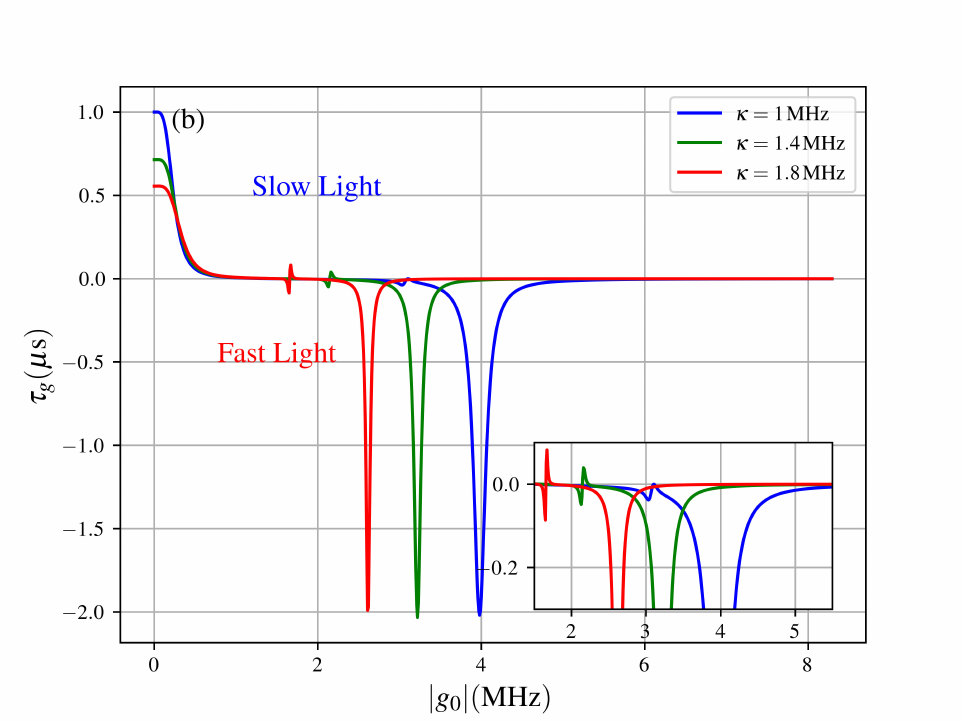}
	\caption{The group delay ($\tau_g$) versus the optomechanical coupling $\vert g_{0} \vert$, for (a) the single-mechanical-mode configuration and (b) the  two-mechanical-modes configuration, at various values of the cavity damping rate $\kappa$. The parameters used in panels (a) and (b) are, respectively, the same as those in Figs. \ref{fig2} and \ref{fig4}.}
	\label{fig7}
\end{figure}
The fast light is represented by a negative group delay, $\tau_g<0$, and the slow light is described by a positive group delay, $\tau_g>0$. To gain insight into these behaviors, we first plot numerical result for the phase $ \Phi_t(\omega_p) $ as a function of the normalized optomechanical coupling $\vert g_{0} \vert/\kappa$ and the normalized detuning $\delta/\omega_m$ ($\delta/\omega_{m_1}$) for single-mechanical-mode configuration (two-mechanical-modes configuration) in Fig. \ref{fig6}(a) (Fig. \ref{fig6}(b)). Clearly, for weak optomechanical coupling, the system exhibits the normal phase dispersion resulting in a positive group delay, corresponding to the generation of the slow light. Intriguingly, the anomalous dispersion of the probe field phase appears in both configurations by increasing the coupling strength, which is corresponding to the fast light. As is clear, the behavior of two configuration are different and these findings demonstrate the system’s potential for switching between slow and fast light. 
In Figs. \ref{fig7}(a) and (b), we have plotted the group delay, $\tau_g$, versus the optomechanical coupling, for the single-mechanical-mode and two-mechanical-modes configurations, respectively, at various values of the damping rate of the microwave cavity field, $\kappa$. Figure \ref{fig7}(a) illustrates that for $\vert g_0 \vert < 2 \rm{MHz} $, the group delay is positive corresponding to the generation of slow light. Furthermore, it shows that for suitable values of $g_0$, it is possible to switch the group delay from superluminal to subluminal propagation and vice versa. In Fig. \ref{fig7}(b), the group delay, $\tau_g$, reveals different behavior due to the OMIA effect, exhibiting two switches between superluminal and subluminal propagation, so that allowing for more manipulation of light propagation. In addition, in both setups, the maximum group delay, $\tau_g$, is inversely related to the cavity damping rate, $\kappa$, i.e.,  the larger the cavity damping rate, the shorter the group delay becomes. It can be seen that from $\kappa=1.8\, \rm{MHz}$ to $1\, \rm{MHz}$, the maximum value of the group delay increases from $\tau_g\simeq 0.5\, \rm{\mu s} $ (red line) to $\tau_g=1 \, \rm{\mu s}$ (blue line). As expected, further reduction of the cavity damping rate enhances the group delay, emphasizing the possibility of achieving a larger group delay in an ultrahigh-Q optomechanical circuit system.
Another significant result of our study, is illustrated in Figs. \ref{fig8}(a) and (b), where we have plotted the group delay, $\tau_g$, versus the optomechanical coupling, for both configurations, at various values of the control field power, $P_c$. These figures clearly show that in the second configuration, adjusting the power of control laser allows for more effective manipulation of the slow and fast light compared to the  first configuration. In Fig. \ref{fig8}(b), it is illustrated that as the power of the control laser increases, less coupling strength is needed to switch between slow and fast light. In other words, as we expect, the additional mechanical mode provides more controllable parameters so that the two-mechanical-modes configuration is a promising candidate for a more efficient manipulation of the light propagation. 
\begin{figure}
      \includegraphics[width=0.45\textwidth]{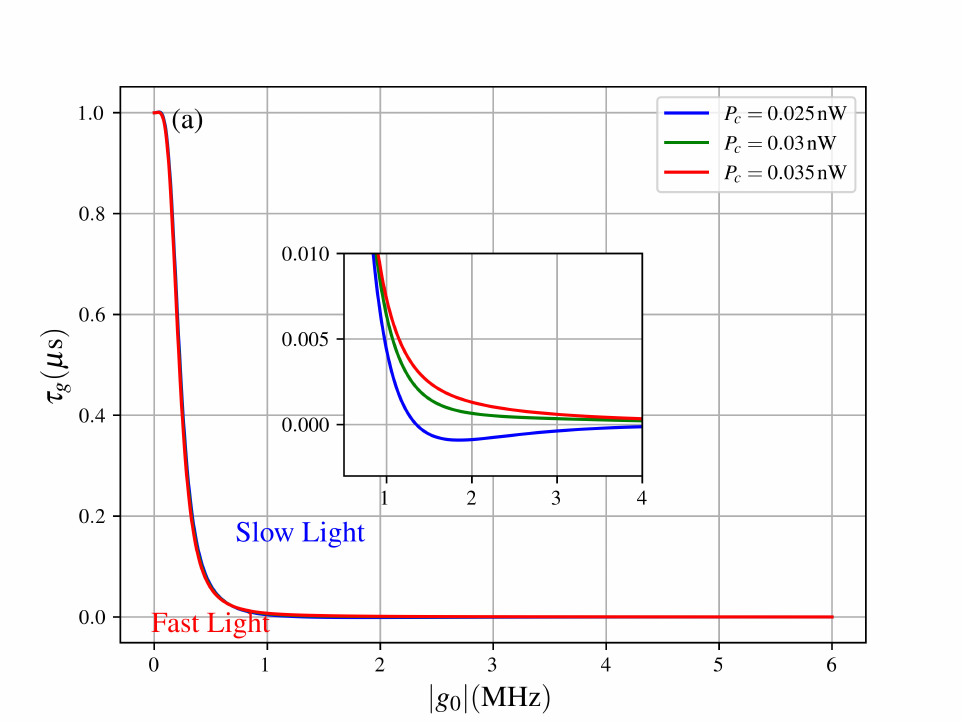}
	\hspace{8mm}
       \includegraphics[width=0.45\textwidth]{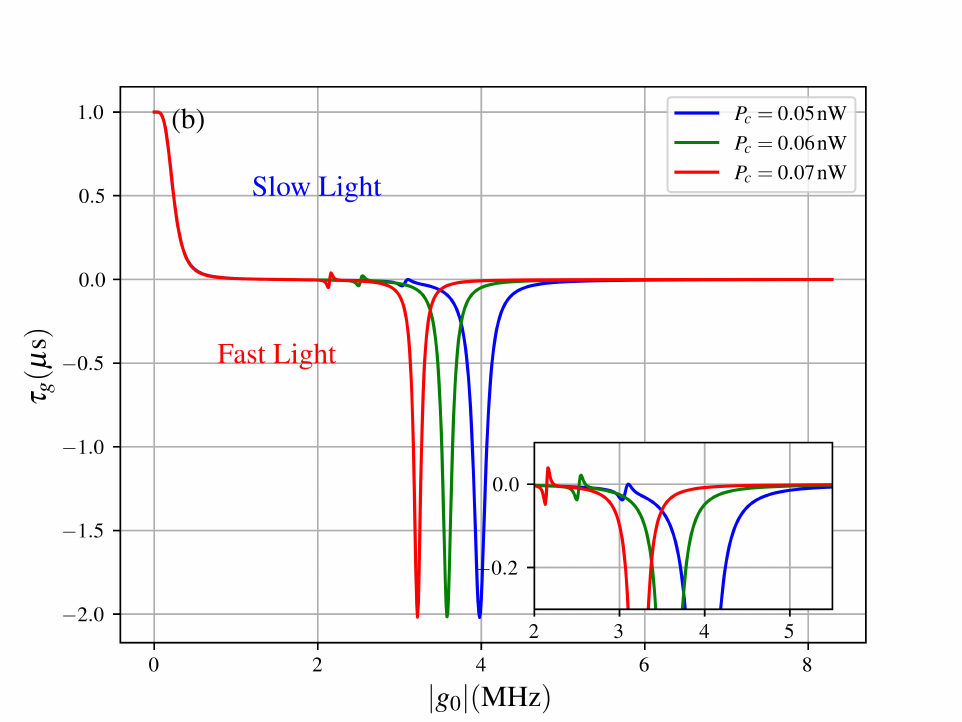}
	\caption{The group delay ($\tau_g$) versus the optomechanical coupling $\vert g_{0} \vert$, for (a) the single-mechanical-mode configuration and (b) the  two-mechanical-modes configuration, at various values of the power of control field $P_c$. The parameters used in panels (a) and (b) are, respectively, the same as those in Figs. \ref{fig2} and \ref{fig4}.}
	\label{fig8}
\end{figure}

\section{\label{sec5} Conclusion and outlook} 
In the present work, we have proposed a microwave-optomechanical circuit to realize the controllable OMIT and OMIA phenomena as well as the tunable conversion between slow and fast light behaviors. The system under consideration, under certain experimentally accessible conditions, exhibits various nonlinear couplings, including the conventional photon-phonon CK, a higher-order generalized phonon-photon CK, a three-mode photon-phonon CK, and phonon-phonon CK couplings. We have shown that the absorption and dispersion properties of the output probe field are significantly sensitive to these nonlinear couplings. In particular, the nonlinear couplings not only can increase the width of the transparency window, but can also enhance the gain and contribute to the amplification of the output probe field in specific frequency regions. Furthermore, we have discussed the group delays for slow and fast light propagation. The results reveal the system’s potential for tunable switching between slow and fast light behaviors.

The obtained results of our study provide new insights into quantum communication and information processing. In addition, the proposed optomechanical circuit is a suitable platform for further investigations concerning quantum state engineering and the mechanical ground-state cooling.
\appendix
\renewcommand{\theequation}{\thesection\arabic{equation}}
\section{DERIVATION OF THE OPTOMECHANICAL HAMILTONIAN (1)}\label{sec6}
In the following, we derive the Hamiltonian of Eq. (\ref{0}). To do this, we start with the Hamiltonian of the proposed microwave-optomechanical system shown in Fig. \ref{fig1}(a): 
\begin{eqnarray}\label{1app}
	\hat{H}_t&=&\hat{H}_{\text{SCPT}_1}+\hat{H}_{\text{SCPT}_2}+\hbar \omega_c^0 \hat{c}^\dagger \hat{c} + \hbar \omega_{m_1}^0 \hat{d}_1^\dagger \hat{d}_1\\ \nonumber
	&& +\hbar \omega_{m_2}^0 \hat{d}_2^\dagger \hat{d}_2
	+ g_{m_1} \hat{\sigma}_3 \hat{x}_{m_1}+ g_{m_2} \hat{\sigma}_3 \hat{x}_{m_2}\\ \nonumber
	&&+ \left( g_{q_1} \hat{\sigma}_1 + g_{q_2} \hat{\sigma}_2 \right) \hat{x}_c^2+  \left( g_{q_3} \hat{\sigma}_1+g_{q_4} \hat{\sigma}_2 \right) \hat{x}_c^2\\ \nonumber
	&& + \left( g_{c_1} \hat{\sigma}_1+g_{c_2} \hat{\sigma}_2\right) \hat{x}_c+ \left( g_{c_3} \hat{\sigma}_1+g_{c_4} \hat{\sigma}_2\right) \hat{x}_c,
\end{eqnarray}
with
\renewcommand{\theequation}{A2\alph{equation}}
\setcounter{equation}{0}
\begin{eqnarray}\label{2app}
	\hat{H}_{\text{SCPT}_1}&=&\frac{B_1}{2} \hat{\sigma}_1+\frac{B_2}{2} \hat{\sigma}_2+\frac{B_3}{2} \hat{\sigma}_3, \\ 
	\hat{H}_{\text{SCPT}_2}&=&\frac{B_1^\prime}{2} \hat{\sigma}_1+\frac{B_2^\prime}{2} \hat{\sigma}_2+\frac{B_3^\prime}{2} \hat{\sigma}_3, 
\end{eqnarray}
\renewcommand{\theequation}{\thesection\arabic{equation}}
and
\renewcommand{\theequation}{A3\alph{equation}}
\setcounter{equation}{0}
\begin{eqnarray}\label{3app}
	g_{m_1}&=&\frac{-4E_c x_{zp,1} (\partial_x C_{g_1}) V_{g_1}}{2e}, \\ 
	g_{m_2}&=&\frac{-4E_c^\prime x_{zp,2} (\partial_x C_{g_2}) V_{g_2}}{2e},\\ 
	g_{q_{1(3)}}&=&\frac{e^2Z_0}{8\hbar}\left(E_{J_{1(3)}}+E_{J_{2(4)}}\right)\cos(\phi/2),\\
	g_{q_{2(4)}}&=&\frac{e^2Z_0}{8\hbar}\left(E_{J_{2(4)}}-E_{J_{1(3)}}\right)\sin(\phi/2),\\ 
	g_{c_{1(3)}}&=&\sqrt{\frac{e^2Z_0}{8\hbar}}\left(E_{J_{1(3)}}+E_{J_{2(4)}}\right)\sin(\phi/2),\\
	g_{c_{2(4)}}&=&\sqrt{\frac{e^2Z_0}{8\hbar}}\left(E_{J_{2(4)}}-E_{J_{1(3)}}\right)\cos(\phi/2).
\end{eqnarray}
\renewcommand{\theequation}{\thesection\arabic{equation}}
This Hamiltonian is a generalization of the Hamiltonian system previously considered in Ref. \cite{96n} to the case in which  two \text{SCPTs} are coupled to a common microwave $LC$ and two independent micromechanical resonators.
Here, the quantities $\omega_c^0=1/\sqrt{LC}$ and $\omega_{m_k}^0$ are the natural frequencies of the cavity and the mechanical modes, respectively. Also, $\hat{x}_{m_{k}}=\hat{d}_{k}^\dagger + \hat{d}_{k}\,(k=1,2)$ and $\hat{x}_c=\hat{c}^\dagger + \hat{c}$ are the respective position operators for the $k$th th mechanical mode and the cavity mode. $\hat{\sigma}_{i} \, \left(i=1,2,3\right)$ stand for the Pauli matrices of the qubits, and $E_{J_i}\,\left(i = 1, 2,3,4\right)$ are the Josephson energies. The effective magnetic fields $B_j$ and $B_j^\prime$ $(j=1,2,3)$ are given by
\renewcommand{\theequation}{A4\alph{equation}}
\setcounter{equation}{0}
\begin{eqnarray}\label{4app}
	B_1&=&-\left(E_{J_1}+E_{J_2}\right) \cos(\phi/2), \\
	B_1^\prime&=&-\left(E_{J_3}+E_{J_4}\right) \cos(\phi/2), \\ 
	B_2&=&\left(E_{J_1}-E_{J_2}\right) \sin(\phi/2), \\
	B_2^\prime&=&\left(E_{J_3}-E_{J_4}\right) \sin(\phi/2), \\
	B_3&=&4 E_c\left(1+\delta n_{g_{0,1}}\right), \\ 
	B_3^\prime&=&4 E_c^\prime\left(1+\delta n_{g_{0,2}}\right), 
\end{eqnarray}
\renewcommand{\theequation}{\thesection\arabic{equation}}
 where $\phi/2$ is the average phase difference of the superconducting order parameters across the junction. Also, $g_{m_{1(2)}}$, $g_{q_{1(3)}}$, $g_{q_{2(4)}}$, $g_{c_{1(3)}}$ and $g_{c_{2(4)}}$ are the coupling strengths between the qubit and resonators. Besides, $e$ is the electric charge unit, and $Z_0= \sqrt{L/C}$ with $L$ and $C$ being the geometric inductance and capacitance \cite{115n}. For simplicity, we suppose $E_{J_1}=E_{J_2}=E_{J}$, $E_{J_3}=E_{J_4}=E_{J}^\prime$
and $\phi=0$. Therefore, $g_{q_2}=g_{q_4}=g_{c_{1(2)}}=g_{c_{3(4)}}=0$, $B_2=B_2^\prime=0$, $g_{q_1}\equiv g_q$ and $g_{q_3}\equiv g_q^\prime$. Then, we have
\begin{eqnarray}\label{5app}
	\hat{H}_t&=&\left(\frac{B_1+B_1^\prime}{2}\right) \hat{\sigma}_1+\left(\frac{B_3+B_3^\prime}{2}\right) \hat{\sigma}_3\\ \nonumber
	&&	+\hbar \omega_c^0 \hat{c}^\dagger \hat{c} + \hbar \omega_{m_1}^0 \hat{d}_1^\dagger \hat{d}_1 +\hbar \omega_{m_2}^0 \hat{d}_2^\dagger \hat{d}_2\\ \nonumber
	&&+\left( g_{m_1} \hat{x}_{m_1}+ g_{m_2}  \hat{x}_{m_2} \right) \hat{\sigma}_3
	+ \left( g_{q_1} + g_{q_1}^\prime \right)  \hat{x}_c^2 \hat{\sigma}_1.
\end{eqnarray} 

In the dispersive limit, i.e., $\hbar \omega_{c,m_1}^{(0)} \ll \vert B_0  \vert = \sqrt{B_{1}^{2}+B_3^{2}}$ and $\hbar \omega_{c,m_2}^{(0)} \ll \vert B_0 ^{\prime} \vert = \sqrt{(B_{1}^\prime)^2+(B_3^\prime)^2}$,  together with the assumption that all couplings are small enough, it is sufficient to diagonalize 
only the interaction part of the Hamiltonian (\ref{5app}) in the qubit basis. Therefore, it can be written as
\begin{eqnarray}\label{6app}
	\hat{H}_t= \begin{pmatrix}
		\frac{B_3+B_3^\prime}{2}+ g_{m_1} \hat{x}_{m_1}+ g_{m_2}  \hat{x}_{m_2}  &\frac{B_1+B_1^\prime}{2}+ g_q \hat{x}_c^2+ g_q^\prime  \hat{x}_c^2 \\		
		\frac{B_1+B_1^\prime}{2}+ g_q \hat{x}_c^2+ g_q^\prime  \hat{x}_c^2  &-\frac{B_3+B_3^\prime}{2}- g_{m_1} \hat{x}_{m_1}- g_{m_2}  \hat{x}_{m_2} 
	\end{pmatrix}, \nonumber\\
\end{eqnarray}
If we assume that both qubits are in their ground states, we can use the replacement $\hat{\sigma}_3 \to -1$. So, corresponding eigenvalues of $\hat{H}_t$ are given by
\begin{eqnarray}\label{7app}
	\lambda=-\frac{\tilde{B}}{2}(1+x)^{1/2},
\end{eqnarray}
where $\tilde{B}^2\equiv \left(B_1+B_1^\prime \right)^2+\left(B_3+B_3^\prime\right)^2$ with
\begin{eqnarray}\label{8app}
	x&=&\frac{4}{\tilde{B}}\left( (B_3+B_3^\prime)( g_{m_1} \hat{x}_{m_1}+ g_{m_2}  \hat{x}_{m_2})+(B_1+B_1^\prime)(g_q+g_q^\prime)\hat{x}_c^2\right)\nonumber\\ 
	&&+\frac{4}{\tilde{B}}\left(( g_{m_1} \hat{x}_{m_1}+ g_{m_2}  \hat{x}_{m_2})^2+(g_q+g_q^\prime)^2 \hat{x}_c^4 \right),
\end{eqnarray}
Assuming $g_{m_1}=-g_{m_2}=g_m$, and writing 
\begin{eqnarray}\label{9app}
	(1+x)^{1/2} \approx 1+\frac{1}{2}x-\frac{1}{8}x^2+\frac{3}{48}x^3-\frac{15}{384}x^4,
\end{eqnarray}
the use of the multinomial theorem \cite{94} together with Eqs. (\ref{7app}) and (\ref{9app}) leads to
\begin{eqnarray}\label{10app}
	\hat{H}_t&=&\hbar \omega_c^0 \hat{c}^\dagger \hat{c} + \hbar \omega_{m_1}^0 \hat{d}_1^\dagger \hat{d}_1 +\hbar \omega_{m_2}^0 \hat{d}_2^\dagger \hat{d}_2\\ \nonumber
	&&+\alpha_m \left(\hat{x}_{m_1}-\hat{x}_{m_2}\right)+\hbar g_\text{Sc}\hat{x}_c^2+\hbar g_\text{Sm}\left(\hat{x}_{m_1}-\hat{x}_{m_2}\right)^2\nonumber\\
	&&+\hbar g_\text{rp} \hat{x}_c^2\left(\hat{x}_{m_1}-\hat{x}_{m_2}\right)+\hbar g_{\text{CK}}^0 \hat{x}_c^2\left(\hat{x}_{m_1}-\hat{x}_{m_2}\right)^2\nonumber\\
	&&+\hbar g_\text{cub}^0 \hat{x}_c^2\left(\hat{x}_{m_1}-\hat{x}_{m_2}\right)^3+\hbar g_\text{quartic}^0 \hat{x}_c^2\left(\hat{x}_{m_1}-\hat{x}_{m_2}\right)^4\nonumber\\
	&&+\hbar G_1^0 \hat{x}_c^4\left(\hat{x}_{m_1}-\hat{x}_{m_2}\right)+\hbar G_2^0 \hat{x}_c^4\left(\hat{x}_{m_1}-\hat{x}_{m_2}\right)^2\nonumber\\
	&&+\hbar G_3^0 \hat{x}_c^4\left(\hat{x}_{m_1}-\hat{x}_{m_2}\right)^3+\hbar G_4^0 \hat{x}_c^4\left(\hat{x}_{m_1}-\hat{x}_{m_2}\right)^4,\nonumber
\end{eqnarray}
where
\begin{eqnarray}\label{11app}
	\alpha&\equiv& -\frac{g_m}{\tilde{B}}\left(B_3+B_3^\prime\right),\\ 
	\hbar g_\text{Sc}&\equiv& -\frac{(g_q+g_q^\prime)}{\tilde{B}}\left(B_1+B_1^\prime\right),\nonumber\\ 
	\hbar g_\text{Sm}&\equiv& -\frac{g_m^2}{\tilde{B}^3}\left(B_1+B_1^\prime\right)^2,\nonumber\\ 
	\hbar g_\text{rp}&\equiv&2\frac{g_m (g_q+g_q^\prime)}{\tilde{B}^3}\left(B_1+B_1^\prime\right)\left(B_3+B_3^\prime\right),\nonumber\\ 
	\hbar g_\text{\text{CK}}^0&\equiv& 2\frac{g_m^2(g_q+g_q^\prime)}{\tilde{B}^5}\left(B_1+B_1^\prime\right)\left(B^2-3(B_3+B_3^\prime)^2\right),\nonumber\\
	\hbar g_\text{cub}^0&\equiv& \frac{4g_m^3(g_q+g_q^\prime)}{\tilde{B}^7}\left(B_1+B_1^\prime\right)\nonumber\\ 
	&&\times \left(B_3+B_3^\prime\right)\left(5(B_3+B_3^\prime)^2-3\tilde{B}^2\right),\nonumber \\
	\hbar g_\text{quartic}^0&\equiv& 2\frac{g_m^4(g_q+g_q^\prime)}{\tilde{B}^9}\left(B_1+B_1^\prime\right)\nonumber\\ 
	&&\times
	\left(-3\tilde{B}^4+30\tilde{B}^2(B_3+B_3^\prime)^2-35(B_3+B_3^\prime)^4\right),\nonumber\\ 
	\hbar G_1^0&\equiv& 2\frac{g_m(g_q+g_q^\prime)^2}{\tilde{B}^5}\left(B_3+B_3^\prime\right)\left(\tilde{B}^2-3(B_3+B_3^\prime)^2\right),\nonumber\\ 
	\hbar G_2^0&\equiv& 2\frac{g_m^2 (g_q+g_q^\prime)^2}{\tilde{B}^7}
	\left(15(B_1+B_1^\prime)^2(B_3+B_3^\prime)^2-2\tilde{B}^4  \right),\nonumber\\ 
	\hbar G_3^0&\equiv& 4\frac{g_m^2 (g_q+g_q^\prime)^2}{\tilde{B}^9}
	\left(B_3+B_3^\prime\right)\nonumber\\ 
	&&\times(5 \tilde{B}^2(B_3+B_3^\prime)^2+15\tilde{B}^2(B_1+B_1^\prime)^2-3\tilde{B}^4\nonumber\\ 
	&&-35(B_1+B_1^\prime)^2(B_3+B_3^\prime)^2  ), \nonumber \\ 
	\hbar G_4^0&\equiv&\frac{g_m^4(g_q+g_q^\prime)^2 }{\tilde{B}^9}\nonumber\\ 
	&&\times
	(60B^2(B_3+B_3^\prime)^2+30\tilde{B}^2(B_1+B_1^\prime)^2-6\tilde{B}^4\nonumber\\ 
	&&-70(B_3+B_3^\prime)^4-420(B_1+B_1^\prime)^2(B_3+B_3^\prime)^2  ),\nonumber
\end{eqnarray}

It should be noted that the Hamiltonian (\ref{10app}) does not contain the linear term in $\hat{x}_c$ due to the assumption $\phi=0$. In addition, in deriving the Hamiltonian in Eq. (\ref{10app}), we have ignored the non-interacting such as $\hat{x}_m^3$ and $\hat{x}_c^4$, because of their negligible contributions compared to the free Hamiltonians of the cavity and mechanical oscillator. Moreover, we consider the interaction terms up to the fourth order in $\hat{x}_c$ and $\hat{x}_m$. The term $\alpha_m (\hat{x}_{m_1}-\hat{x}_{m_2})$ in Eq. (\ref{10app}) describes the qubit-induced static force, which is negligible in our system. The terms with coefficients $g_{Sc}$ and $g_{Sm}$ relate to the cavity and mechanical Stark shifts, respectively. The terms in the third and fourth lines of Eq. (\ref{10app}) represent the radiation-pressure, cross-Kerr, cubic, and quartic couplings. Finally, the terms in the fifth and sixth lines of Eq. (\ref{10app}) are negligible under specific conditions, which we will give in the following.
We now apply the Bogoliubov transformation:
\begin{eqnarray}\label{12app}
	\hat{c}&=&\sinh(\theta_c)\, \hat{a}^\dagger+\cosh(\theta_c)\, \hat{a},\\ \nonumber
	\hat{d}_k&=&\sinh(\theta_{m_k})\, \hat{b}_k^\dagger+\cosh(\theta_{m_k})\, \hat{b}_k,
\end{eqnarray}
where $\theta_{c(m_k)}=(1/2) \ln \left(\omega_{c(m_k)}^0/\sqrt{\omega_{c(m_k)}^0(\omega_{c(m_k)}^0+4g_{\text{Sc}(\text{Sm})})}\right)$  to remove the Stark shift terms from Eq. (\ref{10app}). Besides, provided that 
\begin{eqnarray}\label{13app}
	g_\text{CK}^0,\,g_\text{cub}^0, \,g_\text{quartic}^0, \,G_1^0, \,G_2^0, \,G_3^0, \,G_4^0 \ll g_{0,1(2)},\omega_{m_{1(2)}}, \quad
\end{eqnarray}
we can neglect the terms having unequal powers of the creation and annihilation operators, i.e., $\hat{a}^{\dagger n}\hat{a}^m$  and $\hat{b}^{\dagger n}\hat{b}^m$ with $(n \neq m )$, since they are fast rotating. In the other words, the terms with equal powers of the creation and annihilation operators remain. Therefore, with the above approximations, the effective system Hamiltonian reads as
\begin{eqnarray}\label{14app}
	\hat{H}/\hbar &=&  \omega_a \hat{a}^\dagger \hat{a}+  \omega_{m_k} \hat{b}_k^\dagger \hat{b}_k+ g_{0,k} \hat{a}^\dagger \hat{a} (\hat{b}_k^\dagger+\hat{b}_k)\nonumber\\
	&&+\left(g_{\text{CK},1}+g_{\text{CK},1}^\prime+\tilde g_\text{CK}\right)\,\hat{a}^\dagger \hat{a}\, (\hat{b}_k^\dagger \hat{b}_k)\nonumber\\
	&&+\left(G_{2,k}+G_{4,k}+3G_4\right)\,\hat{a}^\dagger \hat{a} (\hat{b}_k^\dagger \hat{b}_k) \nonumber\\
	&&+\left(g_{\text{CK},k}^\prime+G_{4,k}\right)\hat{a}^\dagger \hat{a} (\hat{b}_k^\dagger \hat{b}_k)^2 \nonumber\\
	&&+\left(2\tilde g_\text{CK}+6G_4\right)(\hat{a}^\dagger \hat{a}) (\hat{b}_1^\dagger \hat{b}_1) (\hat{b}_2^\dagger \hat{b}_2) \nonumber\\
	&&+\left(\tilde g_\text{CK}+G_4\right)(\hat{b}_1^\dagger \hat{b}_1) (\hat{b}_2^\dagger \hat{b}_2),
\end{eqnarray}
where
\begin{subequations}\label{15app}
	\begin{align}
	\omega_c&=\sqrt{\omega_c^0(\omega_c^0+4g_\text{Sc})},\label{eq:A18a}\\
	\omega_{m_k}&=\sqrt{\omega_{m_k}^0(\omega_{m_k}^0+4g_{S{m_k}})},\\
	g_{0,k}&=(-1)^{k+1} 2g_\text{rp}\left(\omega_c^0/\omega_c\right)\left(\omega_{m_k}^0/\omega_{m_k}\right)^{1/2},\\ 
	g_{\text{CK},k}&=4g_{\text{CK}}^0\left(\omega_c^0/\omega_c\right)\left(\omega_{m_k}^0/\omega_{m_k}\right),\\ 
	g_{\text{cub},k}&=(-1)^{k+1}2g_\text{cub}^0\left(\omega_c^0/\omega_c\right)\left(\omega_{m_k}^0/\omega_{m_k}\right)^{3/2},\qquad\\
	g_{\text{cub},k}^\prime&=(-1)^{k+1}6g_{\text{cub}}^0\left(\omega_c^0/\omega_c\right)\left(\omega_{m_k}^0/\omega_{m_k}\right)^{1/2}\left(\omega_{m_{k\pm1}}^0/\omega_{m_{k\pm1}}\right),\qquad\\
	g_{\text{CK},k}^\prime&=12g_{\text{quartic}}^0\left(\omega_c^0/\omega_c\right)\left(\omega_{m_k}^0/\omega_{m_k}\right)^2,\\ 
	\tilde g_{\text{CK}} &=24g_{\text{quartic}}^0\left(\omega_c^0/\omega_c\right)\left(\omega_{m_1}^0/\omega_{m_1}\right)\left(\omega_{m_2}^0/\omega_{m_2}\right),\label{eq:A18h}\\  
	G_{1,k}&=(-1)^{k+1}6G_1^0\left(\omega_c^0/\omega_c\right)^2\left(\omega_{m_k}^0/\omega_{m_k}\right)^{1/2},\\ 
	G_{2,k}&=12G_2^0\left(\omega_c^0/\omega_c\right)^2\left(\omega_{m_k}^0/\omega_{m_k}\right),\\ 
	G_{3,k}&=6G_{3,k}^0\left(\omega_c^0/\omega_c\right)^2\left(\omega_{m_k}^0/\omega_{m_k}\right)^{3/2},\qquad\\
	G_{3,k}^\prime&=(-1)^{k+1}18G_{3,k}^0\left(\omega_c^0/\omega_c\right)^2\left(\omega_{m_k}^0/\omega_{m_k}\right)^{1/2}\left(\omega_{m_{k\pm1}}^0/\omega_{m_{k\pm1}}\right),\qquad\\
	G_{4,k}&=36G_4^0\left(\omega_c^0/\omega_c\right)^2\left(\omega_{m_k}^0/\omega_{m_k}\right)^2,\\ 
	G_{4}&=24G_4^0\left(\omega_c^0/\omega_c\right)^2\left(\omega_{m_1}^0/\omega_{m_1}\right)\left(\omega_{m_2}^0/\omega_{m_2}\right).
	\end{align}
\end{subequations}
The coupling strengths in Eqs. (\ref{11app}) depend on the parameters $B_1, B_1^\prime, B_3, B_3^\prime, g_m, g_m^\prime, g_q, g_q^\prime$. By simplifying, these parameters are given by \cite{115n,96n}:
\begin{subequations}
	\begin{align}
	B_1&=-2E_J, \qquad B_1^\prime=-2E_J^\prime,\\ 
    B_3&=4E_c(1-\delta n_{g_{0,1}}), \\
	B_3^\prime&=4E_c^\prime(1-\delta n_{g_{0,2}}),\\ 
	g_m&\simeq-80 \frac{E_c V_{g_1}C_1}{e\omega_c}, \\
    g_m^\prime&\simeq-80 \frac{E_c^\prime V_{g_2}C_2}{e\omega_c},\\ 
	g_q&=\frac{e^2\,Z_0}{4\hbar}E_J, \\	g_q^\prime&=\frac{e^2\,Z_0}{4\hbar}E_J^\prime,	
	\end{align}
\end{subequations}
As detailed in Ref \cite{115n}, the experimental parameters such as $E_J, E_J^\prime , V_{g_{1(2)}}, C_{1(2)}, Z_0, \delta n_{g_{0,1(2)}}, E_J/E_c$,  and $E_J^\prime/E_c^\prime$ can be used as control parameters for adjusting the optomechanical and nonlinear \text{CK} coupling strengths in Hamiltonian (\ref{14app}). Considering the case
\begin{eqnarray}
G_{2,k}, G_{4,k},G_4, G_4^\prime \ll g_{0,1(2)}, g_{\text{CK},1(2)}, g_{\text{CK},1(2)}^\prime,
\end{eqnarray}
we arrive at the final form of the effective Hamiltonian as
\begin{eqnarray}
	\hat{H}/\hbar &=&  \omega_a \hat{a}^\dagger \hat{a}+  \omega_{m_k} \hat{b}_k^\dagger \hat{b}_k+ g_{0,k} \hat{a}^\dagger \hat{a} (\hat{b}_k^\dagger+\hat{b}_k)+\bar g_{\text{CK},k}\hat{a}^\dagger \hat{a} (\hat{b}_k^\dagger \hat{b}_k) \nonumber\\
	&&+g_{\text{CK},k}^\prime\hat{a}^\dagger \hat{a} (\hat{b}_k^\dagger \hat{b}_k)^2+\tilde{g}_{\text{CK}}(2\hat{a}^\dagger \hat{a}+1) (\hat{b}_1^\dagger \hat{b}_1) (\hat{b}_2^\dagger \hat{b}_2),\quad
\end{eqnarray}
\renewcommand{\theequation}{\thesection\arabic{equation}}
where $\bar{g}_{\text{CK},k}=g_{\text{CK},k}+g_{\text{CK},k}^\prime+\tilde{g}_{\text{CK}}$ is the modified \text{CK} coupling and $\tilde{g}_{\text{CK}}$ denotes the three-mode \text{CK} coupling.  

\section{ DERIVATION OF EQUATIONS (\ref{11c}) AND (\ref{12c})}\label{sec8}
According to Eq. (\ref{121}), the expression for the total output field, $\varepsilon_t$, is given by
\begin{eqnarray}\label{1cc}
	\varepsilon_t= \frac{2 \kappa\,(1+i\,f(\delta)) }{\kappa-i(\delta-\tilde{\Delta}_a)-2\tilde{\Delta}_a\,f(\delta)},
\end{eqnarray}
with
\begin{eqnarray}\label{2cc}
	f(\delta)&=&\frac{2\tilde \Delta_m g_{\text{eff}}^2}{\left((\gamma-i\delta)^2+\tilde \Omega_m^2\right)\left(\kappa-i(\delta+\tilde\Delta_a)\right)},\\ 
	\tilde{\Delta}_a&=&\Delta_a+g_{0} (b_{0}+b_{0}^*)+\bar{g}_{\text{CK}}\, \vert b_{0} \vert ^2+g^\prime_{\text{CK}}\, \vert b_{0} \vert^4, \\
		\tilde \Omega_m^2 &\equiv& \tilde \Delta_m^2+2g_{11}\tilde\Delta_m,\\
	\tilde{\Delta}_{m}&=&\omega_{m}+\bar{g}_{\text{CK}}\, \vert a_0 \vert^2 +2\,g^\prime_{\text{CK}}\, \vert a_0 \vert^2\,\vert b_{0} \vert^2.
\end{eqnarray}
Under the approximations $\tilde \Delta_a \simeq \tilde\Omega_m\equiv\Delta$, $\kappa,\gamma\ll\Delta$ (resolved-sideband limit $\kappa\ll\omega_m$), and $\Delta^2-\delta^2+\gamma^2-i\gamma \delta+\tilde\Omega_m^2\simeq-2\Delta(x+i\gamma)$ with $x=\delta-\Delta$, 
we can rewrite Eq. (\ref{1cc}) as
\begin{eqnarray}\label{4cc}
	\varepsilon_t\simeq \frac{2 \kappa }{\kappa-ix+\frac{g_\text{eff}^2}{\gamma-ix}}.
\end{eqnarray}
The transmission of the probe field is defined as \cite{31n} 
\begin{eqnarray}\label{5cc}
	t_p&=&1-\frac{2 \kappa \vert A_-\vert }{\varepsilon_p}.
\end{eqnarray}
To simplify Eq. (\ref{5cc}), we apply the approximation $\kappa - ix \simeq \kappa$ in Eq. (\ref{4cc}), which yields
\begin{eqnarray}\label{6cc}
	t_p&=&1-2+\frac{2 g_{\text{eff}}^2}{\kappa(\gamma-ix)+g_{\text{eff}}^2}.
\end{eqnarray}
To clearly see the effects of OMIT, it is convenient to introduce the normalized transmission of the probe \cite{31n}:
\begin{eqnarray}\label{7cc}
	t_p^\prime\equiv\frac{t_p-t_r}{1-t_r},
\end{eqnarray}
where 
\begin{eqnarray}\label{8cc}
	t_r&=&t_p(x=0,g_{\text{eff}}=0)=-1.
\end{eqnarray}
Therefore we obtain
\begin{eqnarray}\label{9cc}
	t_p^\prime=\frac{g_{\text{eff}}^2}{\kappa(1-ix)+g_{\text{eff}}^2}.
\end{eqnarray}
The optomechanically induced transparency window is thus given by
\begin{eqnarray}\label{10cc}
	\vert t_p^\prime \vert^2=\frac{g_{\text{eff}}^4/\kappa^2}{\left(\gamma+g_{\text{eff}}^2/\kappa\right)^2+x^2}.
\end{eqnarray}
This is a Lorentzain, peaked at $x=0\,(\delta=\Delta)$, whose width is given by
\begin{eqnarray}\label{11cc}
	\Gamma_{\text{OMIT}}&=&\gamma+\frac{g_{\text{eff}}^2}{\kappa}.
\end{eqnarray}
On the other hand, the imaginary part $\text{Im}\left[{\varepsilon_t} \right]=\varepsilon_i $ which describes the dispersive behavior of the system to the probe field is obtained as:
	\begin{eqnarray}\label{12cc}
		\text{Im}\left[ \varepsilon_t\right]=\frac{-2xg_{\text{eff}}^2/\kappa}{\left(\gamma+g_{\text{eff}}^2/\kappa\right)^2+x^2},
	\end{eqnarray}
The dispersion curve slope can be obtained as
	\begin{eqnarray}\label{13cc}
		\text{K}=\frac{\partial \text{Im}\left[\varepsilon_t\right]}{\partial x}=\frac{\left(-2g_{\text{eff}}^2/\kappa\right)\left((\gamma+g_{\text{eff}}^2/\kappa)^2-x^2 \right)}{\left((\gamma+g_{\text{eff}}^2/\kappa)^2+x^2\right)^2},
	\end{eqnarray}
The dispersion curve slope will take the maximum  $\text{K}_{\text{max}}$ at the transparency window where $x=0$, so we have
	\begin{eqnarray}\label{14cc}
		\text{K}_{\text{max}}=\frac{-2g_{\text{eff}}^2/\kappa}{\left(\gamma+g_{\text{eff}}^2/\kappa\right)^2}.
	\end{eqnarray}
Finally, from Eqs. (\ref{11cc}) and (\ref{14cc}) we get 
	\begin{eqnarray}\label{15cc}
		\Gamma_{\text{OMIT}}\times \text{K}_{\text{max}}&=& \left(\gamma+\frac{g_{\text{eff}}^2}{\kappa}\right)\left(\frac{-2g_{\text{eff}}^2/\kappa}{(\gamma+g_{\text{eff}}^2/\kappa)^2}\right)\\ \nonumber
		&&=\left(\frac{-2g_{\text{eff}}^2/\kappa}{\gamma+g_{\text{eff}}^2/\kappa}\right)\\ \nonumber
		&&=\frac{-2}{\gamma \kappa/g_{\text{eff}}^2+1}.
	\end{eqnarray}

\section{DERIVATION OF EQUATIONS (\ref{8aa}) AND (\ref{8aaa})}\label{sec9}
For the case of the two-mechanical-mode optomechanical system, shown in Fig.\ref{fig1}(b), one can determine the width of the transparency window as well as the slope of the dispersion curve by using Eqs.(\ref{44}), (\ref{45}), and (\ref{12}). We apply the same approximations as in Appendix \ref{sec8}. Considering the resolved-sideband limit  $\kappa\ll\omega_{m_k}$ together with the  approximations $\kappa - i x \simeq \kappa$, $\tilde \Delta_a \simeq \tilde \Delta_{m_k}\simeq\tilde\Omega_{m_k}\equiv\Delta$,  
the response of the output probe field $\varepsilon_t$, reads as:
\begin{eqnarray}\label{4ca}
	\varepsilon_t\simeq \frac{2 \kappa }{\kappa+\frac{G_\text{eff}^2}{\gamma-iX}},
\end{eqnarray}
where $G_\text{eff}^2\equiv g_\text{eff,1}^2+g_\text{eff,2}^2$ and $X\equiv x-g_{mm}\left(d_{12}+d_{21}\right)$ with $d_{12}=\frac{g_{\text{eff},1}}{g_{\text{eff},2}}$ and $d_{12}=\frac{g_{\text{eff},2}}{g_{\text{eff},1}}$. Since Eq. (\ref{4ca}) resembles Eq. (\ref{4cc}) one can follow the same approach as in Appendix \ref{sec8} to find $\Gamma_{\text{OMIT}}$ and $\text{K}_{\text{max}}$. The corresponding Lorentzian linewidth is obtained as:
\begin{eqnarray}\label{5ca}
	\Gamma_{\text{OMIT}}&=&\gamma+\frac{g_{\text{eff,1}}^2+g_{\text{eff,2}}^2}{\kappa}.
\end{eqnarray}
In contrast to the case of the single-mechanical-mode optomechanical system, in which the transparency window is centered at 
$x = 0$,  in the case of the two-mechanical-mode optomechanical system the center is shifted to a nonzero value given by $x=g_{mm}\left(d_{21}+d_{12}\right)=g_{mm}\left(\frac{g_{\text{eff},2}}{g_{\text{eff},1}}+\frac{g_{\text{eff},1}}{g_{\text{eff},2}}\right)$, due to the CK coupling between the two mechanical modes.

On the other hand, we can easily find the imaginary part $\varepsilon_i=\text{Im}\left[{\varepsilon_t} \right]$ as
\begin{eqnarray}\label{6ca}
	\varepsilon_i=\text{Im}\left[ \varepsilon_t\right]=\frac{-2X\left(g_{\text{eff,1}}^2+g_{\text{eff,2}}^2\right)/\kappa}{\left(\gamma+\left(g_{\text{eff,1}}^2+g_{\text{eff,2}}^2\right)/\kappa\right)^2+X^2}.
\end{eqnarray}
The dispersion curve slope can be obtained as
\begin{eqnarray}\label{7ca}
	\text{K}&=&\frac{\partial \text{Im}\left[\varepsilon_t\right]}{\partial X}\\
	&=&\frac{\left(-2\left(g_\text{eff,1}^2+g_\text{eff,2}^2\right)/\kappa\right)\left(\left(\gamma+\left(g_\text{eff,1}^2+g_\text{eff,2}^2\right)/\kappa\right)^2-X^2 \right)}{\left(\left(\gamma+\left(g_\text{eff,1}^2+g_\text{eff,2}^2\right)/\kappa\right)^2+X^2\right)^2}.\nonumber
\end{eqnarray}
The dispersion curve slope will take the maximum  $\text{K}_{\text{max}}$ at the transparency window where $X=0$, so we have
\begin{eqnarray}\label{8ca}
	\text{K}_{\text{max}}=\frac{-2\left(g_{\text{eff,1}}^2+g_{\text{eff,2}}^2\right)/\kappa}{\left(\gamma+\left(g_{\text{eff,1}}^2+g_{\text{eff,2}}^2\right)/\kappa\right)^2}.
\end{eqnarray}
Finally, from Eqs. (\ref{5ca}) and (\ref{8ca}) we get
\begin{eqnarray}\label{9ca}
	\Gamma_{\text{OMIA}}\times \text{K}_{\text{max}}&=& \left(\gamma+\frac{g_{\text{eff,1}}^2+g_{\text{eff,2}}^2}{\kappa}\right)\left(\frac{-2(g_{\text{eff,1}}^2+g_{\text{eff,2}}^2)/\kappa}{\left(\gamma+\left(g_{\text{eff,1}}^2+g_{\text{eff,2}}^2\right)/\kappa\right)^2}\right),\nonumber\\ \nonumber
	&&=\left(\frac{-2\left(g_{\text{eff,1}}^2+g_{\text{eff,2}}^2\right)/\kappa}{\gamma+\left(g_{\text{eff,1}}^2+g_{\text{eff,2}}^2\right)/\kappa}\right),\nonumber\\ 
	&&=\frac{-2}{\left(\frac{\gamma \kappa}{g_{\text{eff,1}}^2+g_{\text{eff,2}}^2}\right)+1}.
\end{eqnarray}

\end{document}